\documentclass[a4paper,11pt]{JHEP3}

\usepackage{amssymb}

\def\bra#1{\left\langle#1\right|}
\def\ket#1{\left|#1\right\rangle}

\author{
Leszek Hadasz\footnote{e-mail: hadasz@th.if.uj.edu.pl} \\
    M. Smoluchowski Institute of Physics,
    Jagiellonian University,
    Reymonta 4,
    30-059~Krak\'ow, Poland}
\author{
Zbigniew Jask\'olski\footnote{jask@ift.uni.wroc.pl}\\
    Institute of Theoretical Physics,
    University of Wroc{\l}aw,
    pl. M. Borna, 50-204~Wroc{\l}aw, Poland
    }
\author{
Paulina Suchanek\footnote{suchanek@th.if.uj.edu.pl} \\
    M. Smoluchowski Institute of Physics,
    Jagiellonian University,
    Reymonta 4, 30-059~Krak\'ow, Poland}

\abstract{The structure of the 4-point $N=1$ super-conformal blocks in the Ramond sector is analyzed.
The elliptic recursion relations for these blocks are derived.}

\title{Elliptic recurrence representation of the N=1 superconformal blocks in the Ramond sector}

\preprint{}

\begin{document}

\section{Introduction}

Correlation functions in any 2-dimensional
CFT can be expressed as sums (or integrals) of three-point coupling
constants and some universal, model independent functions called conformal blocks  \cite{Belavin:1984vu}.
Even in a simple case of the 4-point conformal block its direct calculation
is prohibitively complicated.
Efficient recursive methods of an approximate, analytic calculation
of a general 4-point conformal block has been pioneered long time ago by Al.~Zamolodchikov
\cite{Zamolodchikov:ie,Zamolodchikov:2,Zamolodchikov:3}.
His method was used for instance in checking the conformal bootstrap in the Liouville field theory with the
Dorn, Otto, Zamolodchikov and Zamolodchikov
coupling constants \cite{Zamolodchikov:1995aa}, in a study of the $c\to 1$ limit of minimal models
\cite{Runkel:2001ng} or in obtaining new results in the classical geometry of hyperbolic surfaces
\cite{Hadasz:2005gk}.

Recently a recursion representation has been worked out
for the super-conformal blocks related to the Neveu-Schwarz algebra
\cite{Hadasz:2006sb,Belavin:2007zz,Belavin:2007gz,Belavin:2007eq}. The so called
elliptic recursion was conjectured in \cite{Belavin:2007gz} for one type of NS blocks
and used for a numerical
verification of the consistency of $N =1$ super-Liouville field theory.
An extension of this method to another type of NS blocks was proposed
in \cite{Belavin:2007eq} where also further numerical support for the consistency of the
$N =1$ super-Liouville theory was presented.
A comprehensive derivation
of the elliptic recursion for all types of NS blocks was given in \cite{Hadasz:2007nt}.

In the present paper we address the problem of the elliptic recurrence for
conformal blocks in the Ramond sector of N=1 SCFT. We restrict ourselves to  the class of SCFT models
with the tensor product ${\cal R} \otimes  \bar {\cal R}$ of the left $\cal R$ and the right $\bar {\cal R}$
Ramond algebras extended by the common (for the left and the right sector) parity operator $(-1)^F$. We shall also consider
only  4-point blocks with Ramond external states which correspond to factorization on Neveu-Schwarz states.

There are some features of the Ramond sector which make a standard analysis of
conformal blocks more
difficult than in the NS and the bosonic cases.\footnote{The structure of the 3-point conformal blocks in the Ramond sector
is scarcely discussed in literature \cite{Friedan:1984rv,Gaberdiel:1996kf}. We are not aware of any
discussion on the 4-point blocks with external Ramond states.}
First of all the  structure of 3-point conformal blocks is  more complicated.
Since the correlation functions of one fermionic current, two Ramond and one Neveu-Schwarz fields
are double valued the standard contour deformation
arguments do not work. One way to avoid this problem has been
proposed in the early days of the 2-D SCFT \cite{Friedan:1984rv}
(see also  \cite{Gaberdiel:1996kf} for more detailed analysis).
In the present paper we follow essentially the same approach considering
single valued functions
$$
\langle \phi(\infty,\infty) S(w) R_2(z,\bar z)R_1(0,0)\rangle\sqrt{(w-z)w}
$$
and expressing contour integrals around a location of each field in terms
of fields excitations.
This leads to the Ward identities (\ref{wardid1}), (\ref{wardid2})
which determine the 3-point blocks up to four independent constants.
Fortunately in spite of their complicated form one can derive all properties
of the 3-point blocks required for the derivation of the elliptic recurrence formula.
Let us mention that the so called $x$-expansion seems to be available
in the present case only via the elliptic recurrence.

The second complication arises from the fact
that the tensor product of the left and the right chiral structures
has to be reduced to an irreducible representation
of ${\cal R} \otimes  \bar {\cal R}$
extended by $(-1)^F$.
This reduction is responsible for the reduction of eight
independent constants hidden in 3-point blocks into two independent
structure constants of the Ramond sector. It also determines
the representation of the Ramond fields in terms of chiral vertex operators and suggests
a convenient basis of the 3-point blocks
for which this representation is diagonal.

Once the structure and properties of the 3-point blocks are clarified
the definition of 4-point conformal blocks is straightforward. As in the case of
the NS sector one gets four even and four odd conformal blocks.
With the appropriate definition at hand one can follow Zomolodchikov's method
of the derivation of the elliptic recurrence \cite{Zamolodchikov:3}.
As in the NS case \cite{Hadasz:2007nt} the arguments concerning large intermediate weight $\Delta$
asymptotic are based on the quasiclassical limit
of the path integral representation of the Liouville theory.
Regular terms of elliptic blocks can be  calculated from
the  $\hat c=1$ conformal block with $\Delta_i=\frac{1}{16}$  Ramond external states
 and an arbitrary intermediate weight.
An explicit formula for this blocks can be obtained using the techniques
of the chiral superscalar model \cite{Hadasz:2007ns}.

The organization of the paper is as follows. In Section 2 we present our
notation and basic properties of the 3-point functions in the Ramond sector \cite{Zamolodchikov:1988nm}.
Section 3 is devoted to the basic structure of the highest weight chiral module for the Ramond algebra extended by the chiral
parity operator \cite{Dorrzapf:1999nr}. The reduction of the tensor product
of the left and the right modules to an irreducible representation of  ${\cal R} \otimes  \bar {\cal R}$
extended by $(-1)^F$ is briefly described.
In Section 4 we use the Ward identities to determine the properties of the 3-point conformal blocks.
This section contains the main results of the present paper and paves a way for
an appropriate definition of the conformal blocks.
In Section 5 we define 4-point conformal blocks and analyze their analytic properties as functions
of the intermediate weight. The main result of this section is a calculation of the residua at singular weights.
As a side result we obtained a universal property of the Ramond structure constants $C^\pm$ (\ref{constant1})
in a general
$N=1$ SCFT. If the even fusion rules (\ref{efusionRules}) are satisfied
$
C^{+} =
- C^{-}
$, while for the odd fusion rules (\ref{ofusionRules}) one has $
C^{+} =
 C^{-}.
$
In Section 6 we discus the large $\Delta$ asymptotic of the conformal blocks and derive the elliptic recurrence formula.
The regular terms of elliptic blocks are also calculated.

There are some problems which are natural continuation of the present work.
First of all one should extend the analysis to  the
blocks related to the factorization on Ramond states.
This can be done with the techniques developed in the present paper.
The second possible topic is to extend the elliptic recurrence methods to
the $N=2$ SCFT.
Let us also mention that one can use our results to check consistency of
the $N=1$ super Liouville theory \cite{Poghosian:1996dw} or its $\hat c\to
1$
limit \cite{Fredenhagen:2007tk}.

\section{Three-point correlation functions of the Ramond sector }
The superconformal symmetry is generated by a pair of holomorphic
currents $T(z),\, S(z)$ (and their anti-holomorphic counterparts
${\overline T}(\bar z),\,{\overline S}(\bar z)$) satisfying  the
OPE-s
\begin{eqnarray}
\label{OPE:TS}
\nonumber
T(z)T(0) & = & \frac{c}{2z^4} + \frac{2}{z^2}T(0) + \frac{1}{z}\partial T(0) + \ldots,
\\
T(z)S(0) & = & \frac{3}{2z^2}S(0) + \frac{1}{z}\partial S(0)+ \ldots,
\\
\nonumber
S(z)S(0) & = & \frac{2c}{3z^3} + \frac{2}{z}T(0) + \ldots\,.
\end{eqnarray}
The space of fields  of superconformal field theory (hereafter SCFT) decomposes onto
the space of the Neveu-Schwarz (or NS for short) fields $\phi_{\rm NS}$ local with respect to $S(z),$ and the
space of the Ramond fields $R$ with the property that any correlation function of the form
$
\left\langle
S(z)R(z_1,\bar z_1)\ldots
\right\rangle
$
changes the sing upon analytic continuation
in $z$ around the point $z=z_1.$
This property implies the following form of the OPE:
\[
S(z)R(0,0) \; = \; \sum\limits_{m\in {\mathbb Z} } z^{m-\frac32}S_{-m}R(0,0).
\]
Together with the usual Virasoro generators $L_n$ defined by
\[
T(z)R(0,0) \; = \; \sum\limits_{n\in {\mathbb Z}} z^{n-2}L_{-n}R(0,0),
\]
$S_k$ form the Ramond algebra determined by (\ref{OPE:TS}),
\begin{eqnarray}
\label{R}
\nonumber
\left[L_m,L_n\right] & = & (m-n)L_{m+n} +\frac{c}{12}m\left(m^2-1\right)\delta_{m+n},
\\
\left[L_m,S_n\right] & = &\frac{m-2n}{2}S_{m+n},
\\
\nonumber
\left\{S_m,S_n\right\} & = & 2L_{m+n} + \frac{c}{3}\left(m^2 -\frac14\right)\delta_{m+n}.
\end{eqnarray}

In the space of all R fields there exist ``super-primary'' fields
$R_{\Delta,\bar\Delta}^{\pm}(u,\bar u)$ with the conformal weights
$\Delta, \bar \Delta,$ which satisfy the OPE's
\footnote{Following \cite{Zamolodchikov:1988nm} we chose the ``symmetric''
convention for $\pm$ components of the Ramond fields.}
\begin{eqnarray}
\label{primary:T}
\nonumber
T(z) R_{\Delta,\bar\Delta}^{\pm}(u,\bar u)
& \sim &
{\Delta \over (z-u)^2}R_{\Delta,\bar\Delta}^{\pm}(u,\bar u)
+
{1\over z-u}\partial
R_{\Delta,\bar\Delta}^{\pm}(u,\bar u),
\\[6pt]
\bar T(\bar z) R_{\Delta,\bar\Delta}^{\pm}(u,\bar u)
& \sim &
{\bar \Delta \over (\bar z-\bar u)^2}R_{\Delta,\bar\Delta}^{\pm}(u,\bar u)
+
{1\over \bar z-\bar u}\partial
R_{\Delta,\bar\Delta}^{\pm}(u,\bar u)
\end{eqnarray}
and
\begin{eqnarray}
\label{primary:S}
\nonumber
S(z) R_{\Delta,\bar\Delta}^{\pm}(u,\bar u)
& \sim &
{i \beta {\rm e}^{\mp i{\pi\over 4}} \over (z-u)^{3\over 2}}R_{\Delta,\bar\Delta}^{\mp}(u,\bar u),
\\[6pt]
\bar S(\bar z) R_{\Delta,\bar\Delta}^{\pm}(u,\bar u)
& \sim &
{-i\bar\beta {\rm e}^{\pm i{\pi\over 4}} \over (\bar z-\bar u)^{3\over 2}}R_{\Delta,\bar\Delta}^{\mp}(u,\bar u),
\end{eqnarray}
where $\beta,\bar \beta$ are related to the conformal weights by
$$
\Delta = {c\over 24}-\beta^2,
\hskip 1cm
\bar \Delta = {c\over 24}-\bar \beta^2.
$$

Using projective
 transformations one can express three-point function involving two R primary fields
and one NS primary superfield
$$
\Phi_3(z,\theta;\bar z,\bar\theta)
=
\varphi(z,\bar z) +\theta \psi(z,\bar z) +\bar \theta \bar \psi (z,\bar z)
+i \theta \bar \theta \widetilde \varphi(z,\bar z)
$$
in the form
\begin{eqnarray}
\nonumber
\Big\langle
\Phi_3(z_3,\theta_3;\bar z_3,\bar\theta_3)
R^{\epsilon_2}_2(z_2,\bar z_2)
R^{\epsilon_1}_1(z_1,\bar z_1)
\Big\rangle
&=&
z_{32}^{\gamma_1}\bar z_{32}^{\bar \gamma_1}\,
z_{31}^{\gamma_2}\bar z_{31}^{\bar \gamma_2}\,
z_{21}^{\gamma_3}\bar z_{21}^{\bar \gamma_3}\
\\
&&\hspace{-110pt} \times
\left[
\; \delta_{\epsilon_1,\epsilon_2}
\left(
C^{\epsilon_1}_{321}
+\widetilde C^{\epsilon_1}_{321}
\left|
{z_{31}z_{32}\over z_{12}}
\right|i \theta_3\bar\theta_3
\right)
\right.
\\
\nonumber
&&\hspace{-106pt}
\left.\;
+\;
\delta_{\epsilon_1,-\epsilon_2}
\left(
D^{\epsilon_1}_{321}
\left(
{z_{31}z_{32}\over z_{12}}
\right)^{1\over 2}\theta_3
+\bar D^{\epsilon_1}_{321}
\left(
{\bar z_{31}\bar z_{32}\over \bar z_{12}}
\right)^{1\over 2}\bar\theta_3
\right)
\right]
\end{eqnarray}
where $\gamma_1 = \Delta_1 -\Delta_2 -\Delta_3,\ $  $z_{12} = z_1-z_2$ etc.
and
\begin{eqnarray}
\label{constant1}
C^{\pm}_{321} & = & \Big\langle\varphi_3(\infty,\infty)R_2^\pm(1,1)R^\pm_1(0,0)\Big\rangle,
\\[6pt]
\label{constant2}
\widetilde C^{\pm}_{321} & = & \Big\langle\widetilde \varphi_3(\infty,\infty)R_2^\pm(1,1)R^\pm_1(0,0)\Big\rangle,
\\
\label{constant3}
D^{\pm }_{321} & = & \Big\langle\psi_3(\infty,\infty)R_2^\pm(1,1)R^\mp_1(0,0)\Big\rangle,
\\[6pt]
\label{constant4}
\bar D^{\pm }_{321} & = & \Big\langle\bar \psi_3(\infty,\infty)R_2^\pm(1,1)R^\mp_1(0,0)\Big\rangle.
\end{eqnarray}
Due to the Ward identities
\begin{eqnarray*}
\Big\langle S_{-{1\over 2}}\varphi_3(\infty,\infty)R_2^\epsilon (1,1)R^{\epsilon'}_1(0,0)\Big\rangle &  &
\\&&\hspace{-140pt}=\;
\Big\langle\varphi_3(\infty,\infty)S_0 R_2^{\epsilon}(1,1)R^{\epsilon'}_1(0,0)\Big\rangle
+i\epsilon
\Big\langle\varphi_3(\infty,\infty)R_2^{\epsilon}(1,1)S_0 R^{\epsilon'}_1(0,0)\Big\rangle,
\\[6pt]
\Big\langle \bar S_{-{1\over 2}}\varphi_3(\infty,\infty)R_2^\epsilon (1,1)R^{\epsilon'}_1(0,0)\Big\rangle &  &
\\&&\hspace{-140pt}=\;
\Big\langle\varphi_3(\infty,\infty)\bar S_0 R_2^{\epsilon}(1,1)R^{\epsilon'}_1(0,0)\Big\rangle
-i\epsilon
\Big\langle\varphi_3(\infty,\infty)R_2^{\epsilon}(1,1)\bar S_0 R^{\epsilon'}_1(0,0)\Big\rangle,
\end{eqnarray*}
 only two of these structure constants are independent:
\begin{eqnarray}
\nonumber
\widetilde C^{\pm}_{321}
&=& \mp i\left[ (\bar \beta_1\beta_1 +\bar \beta_2\beta_2) C^\pm_{321}
-(\bar \beta_1\beta_2 +\bar \beta_2\beta_1) C^\mp_{321}\right], \\
\label{strcon}
D^\pm_{321}
&=&
i{\rm e}^{\pm i{\pi\over 4}}
\left[
\beta_2C^\mp_{321}
+
\beta_1 C^\pm_{321}
\right],
\\
\nonumber
\bar D_{321}^\pm&=&
-i{\rm e}^{\mp i{\pi\over 4}}\left[
\bar \beta_2C^\mp_{321}
+
\bar \beta_1 C^\pm_{321}
\right].
\end{eqnarray}

\section{R supermodule}

In SCFT one usually works with the Ramond algebra (\ref{R}) extended by the fermion parity
operator $(-1)^F$:
$$
[(-1)^F, L_m]=\{(-1)^F,S_n\}=0,\hskip 1cm m,n\in \mathbb{Z}.
$$
Let $w^+_{\Delta}$ be the highest weight state with respect to the extended Ramond
algebra (\ref{R})
\begin{equation}
\label{highest}
L_0 w^+_{\Delta}=\Delta w^+_{\Delta}\,,
\hspace{15pt}
(-1)^F  w^+_{\Delta} =  w^+_{\Delta}\,,
\hspace{15pt}
L_m w^+_{\Delta}= S_n w^+_{\Delta} =0,
\hspace{15pt}
m,n\in \mathbb{N},
\end{equation}
where $ \mathbb{N}$ is the set of positive integers.
We denote by ${\cal W}^{f}_\Delta$ the free vector space  generated by all vectors
of the form
\begin{equation}
\label{basis}
 w_{\Delta,KM}
\; = \;
S_{-K} L_{-M} w^+_{\Delta}
\; = \;
S_{-k_i}\ldots S_{-k_1}L_{-m_j}\ldots L_{-m_1} w^+_{\Delta}\,,
\end{equation}
where
 $K = \{k_1,k_2,\ldots,k_i\}\subset \mathbb{N}\cup \{0\} $ and
 $M = \{m_1,m_2,\ldots,m_j\}\subset \mathbb{N}$ are
 arbitrary ordered
 sets of  indices
\[
k_i < \ldots < k_2 < k_1 ,
\hskip 1cm
m_j \leq \ldots \leq m_2 \leq m_1,
\]
such that $
|K|+|M|= k_1+\dots+k_i+m_1+\dots+m_j = {f}
$.

The $\mathbb{Z}$-graded representation of the extended Ramond algebra
determined on the space
$$
{\cal W}_\Delta\
=
\hspace*{-3mm}
\bigoplus\limits_{{f}\in \mathbb{N}\cup\{0\}}
\hspace*{-1mm}
{\cal W}^{f}_\Delta
$$
by  relations
(\ref{R}) and (\ref{highest}) is called the R supermodule
of the highest weight $\Delta$ and the central charge $c$
(in order to simplify the notation we omit the subscript $c$ at $\cal W$).
Each ${\cal W}^{f}_\Delta$ is an eigenspace of $L_0$ with the eigenvalue
$\Delta+{f}$. The space ${\cal W}_\Delta$ has also a natural $
\mathbb{Z}_2$-grading:
$$
{\cal W}_\Delta
\; = \;
{\cal W}^+_\Delta \oplus
{\cal W}^-_\Delta\,,
\hskip 15pt
{\cal W}^+_\Delta
\; = \hskip -5pt
\bigoplus\limits_{f\in \mathbb{N}\cup \{0\}}
{\cal W}^{f+}_\Delta\,,
\hskip 15pt
{\cal W}^-_\Delta
\; = \hskip -5pt
\bigoplus\limits_{f\in \mathbb{N}\cup \{0\}}
{\cal W}^{f-}_\Delta\,,
$$
where ${\cal W}^{f\pm}_\Delta$ are common eigenspaces of the
operators $L_0,(-1)^F$. Note that the subspaces ${\cal W}^{0+}_\Delta,{\cal W}^{0-}_\Delta$
are 1-dimensional except the case $\Delta ={c\over 24}$ where
${\cal W}^{0-}_\Delta=\{0\}$.

A nonzero element ${\chi}\in {\cal W}^{f}_\Delta $ of degree ${f}$
is called a singular vector if
it satisfies the highest weight conditions (\ref{highest}) with
$
L_0{\chi} =(\Delta +{f}){\chi}
$.
It generates its own R supermodule ${\cal W}_{\Delta +{f}}$ which is a
submodule of ${\cal W}_{\Delta}$.

The analysis of singular vectors can be facilitated by introducing a
symmetric bilinear form $\langle
.\,,.\rangle_{c,\Delta}$ on ${\cal W}_{\Delta}$ uniquely determined  by the relations
$\langle w^+_{\Delta}, w^+_{\Delta}\rangle =1$, $\langle w^+_{\Delta},S_0w^+_{\Delta}\rangle =0$ and
$(L_{m})^{\dag}=L_{-m},(S_{k})^{\dag}=S_{-k}$.
It is block-diagonal with respect to the $L_0$- and $(-1)^F$-gradings.
We denote by $B^{\,{f}\pm}_{c,\Delta}$ the matrix of $\langle .\,,.\rangle_{c,\Delta}$
on ${\cal W}_{\Delta}^{\,{f}\pm}$ calculated
in the basis (\ref{basis}):
\begin{equation}
{ \label{matrix}
\left[ B^{\,{f}\pm}_{c,\Delta}\right]_{KM,LN}
= \;
\big\langle  w_{\Delta,KM}, w_{\Delta,LN}\big\rangle_{c,\Delta}.}
\end{equation}
It is nonsingular if and only if the R supermodule ${\cal W}_{\Delta}$ does
not contain singular vectors of degrees $0,1,2,\dots,{f}$.
The formula for the determinant of this matrix was conjectured by
Friedan, Qiu and  Shenker
 \cite{Friedan:1984rv} and proven by Meurman and Rocha-Caridi
\cite{Meurman:1986gr}. For level zero it reads
$$
\det  B^{\,{0}+}_{c,\Delta}= 1,\hskip 1cm \det  B^{\,{0}-}_{c,\Delta}= \Delta - {c\over 24},
$$
and for higher levels
\begin{equation}
\label{Kac}
\det B^{\,{f}\pm}_{c,\Delta}
\; = \;
\left(\Delta - {c\over 24}\right)^{P_R(f)\over 2}\hskip -2mm
\prod\limits_{1\leqslant rs \leqslant 2{f}}
(\Delta-\Delta_{rs})^{P_{R}({f}-{rs\over 2})},
\end{equation}
where  $r,s\in \mathbb{N}$, the sum $r+s$ must be odd and
\begin{eqnarray}
\label{delta:rsR}
\Delta_{rs}(c)
& = &
{1\over 16}-\frac{rs-1}{4} + \frac{1-r^2}{8}b^2 + \frac{1-s^2}{8}\frac{1}{b^2}\,,\hskip 10mm
c=\frac{3}{2} +3\left(b+{1\over b}\right)^2.
\end{eqnarray}
The multiplicity of each zero is given by
$P_{R}({f})= \dim {\cal W}^{f}_\Delta$
and can be read off from the relation
$$
\sum\limits_{{f}=0}^\infty P_{R}({f})q^{f}
\; = \;
\prod\limits_{n=1}^\infty {1+q^{n}\over 1-q^n}.
$$

The tensor product ${\cal W}_\Delta \otimes \bar{\cal W}_{\bar\Delta}$ of the left and the right
R supermodules is defined as a graded tensor product of representations of $\mathbb{Z}_2$-graded algebras.
This provides a representation of the direct sum $R\oplus \bar R$ of  left and right Ramond algebras
extended by the left $(-1)^{F_L}$ and the right $(-1)^{F_R}$ parity operators. We are usually interested
in the extension of $R\otimes \bar R$ by the common parity operator
$$
(-1)^F =(-1)^{F_L}(-1)^{F_R}
$$
and the corresponding $\mathbb{Z}_2$-grading.
For $\Delta, \bar\Delta\neq {c\over 24}$
an appropriate representation can be easily obtained restricting the action of
$R\otimes \bar R$ and $(-1)^F$
to an invariant subspace ${\cal W}_{\Delta,\bar \Delta}
\subset {\cal W}_\Delta \otimes \overline{\cal W}_{\bar\Delta}$ generated by the vectors
\begin{equation}
\label{basis0}
\begin{array}{llllll}
 w^+_{\Delta,\bar\Delta}&=&
 {1\over \sqrt{2}}\left( w^+_\Delta \otimes  w^+_{\bar\Delta} -i\,
 w^-_\Delta \otimes  w^-_{\bar\Delta}\right),
\\
 w^-_{\Delta,\bar\Delta}&=&
 {1\over \sqrt{2}}\left(
 w^+_\Delta \otimes  w^-_{\bar\Delta}\; +\;
 w^-_\Delta \otimes  w^+_{\bar\Delta}\right),
\end{array}
\end{equation}
where
$
w^-_\Delta=  {{\rm e}^{i{\pi\over 4}}\over {i} \beta}S_0  w^+_{\Delta}
$
and
$
w^-_{\bar \Delta}=  {{\rm e}^{-i{\pi\over 4}}\over {- i} \bar \beta}\bar S_0  w^+_{\bar \Delta}.
$
We shall call it a ``small'' representation.

The choice of basis (\ref{basis0}) in the zero level subspace ${\cal W}_{\Delta,\bar \Delta}^0$
corresponds to our choice of the Ramond fields (\ref{primary:S})
\begin{eqnarray}
\label{primarystates}
S_0  w^\pm_{\Delta, \bar\Delta}
& = &
i \beta {\rm e}^{\mp i{\pi\over 4}}  w^\mp_{\Delta, \bar\Delta},
\hskip 1cm
\bar S_0  w^\pm_{\Delta, \bar\Delta}
\;=\;
-i \bar \beta {\rm e}^{\pm i{\pi\over 4}}  w^\mp_{\Delta, \bar\Delta}.
\end{eqnarray}

\section{Three-point conformal blocks}
Super-descendants $\varphi_{\Delta,\bar\Delta}(\xi,\bar\xi |z,\bar z)$
of the super-primary field $\varphi_{\Delta,\bar\Delta}( \nu,\bar \nu |z,\bar z)=\varphi_{\Delta,\bar\Delta}(z,\bar z)$ are defined by
the relations:
\begin{eqnarray*}
\varphi_{\Delta,\bar\Delta}(L_{-m}\xi,\bar\xi |z,\bar z)
&=&
\oint {du\over 2\pi i} (u-z)^{1-m}T(u)\varphi_{\Delta,\bar\Delta}(\xi,\bar\xi |z,\bar z),
\hskip 5mm
m\in\mathbb{N},\\
\varphi_{\Delta,\bar\Delta}(S_{-k}\xi,\bar\xi |z,\bar z)
&=&
\oint {du\over 2\pi i} (u-z)^{{1\over 2}-k}S(u)\varphi_{\Delta,\bar\Delta}(\xi,\bar\xi |z,\bar z),
\hskip 5.5mm
k\in\mathbb{N}-\textstyle{1\over 2},
\end{eqnarray*}
and by analogous formulae for the Ramond  sector
\begin{eqnarray*}
R_{\Delta,\bar\Delta}(L_{-m}\eta,\bar\eta |z,\bar z)
&=&
\oint {du\over 2\pi i} (u-z)^{1-m}T(u)R_{\Delta,\bar\Delta}(\eta,\bar\eta |z,\bar z),
\hskip 5mm
m\in\mathbb{N},\\
R_{\Delta,\bar\Delta}(S_{-k}\eta,\bar\eta |z,\bar z)
&=&
\oint {du\over 2\pi i} (u-z)^{{1\over 2}-k}S(u)R_{\Delta,\bar\Delta}(\eta,\bar\eta |z,\bar z),
\hskip 5.5mm
k\in\mathbb{N}.
\end{eqnarray*}
 Using conformal Ward identities one can express
an arbitrary correlator  of three descendants as a sum of terms
which can be further factorized onto a holomorphic and
anti-holomorphic parts. Consider for instance a correlator of
two arbitrary excited Ramond fields $R^\epsilon(z,\bar
z),R^{\epsilon'}(0,0)$ of definite parities $\epsilon, \epsilon'$
and an arbitrary NS excited field $\varphi(\infty,\infty)$. Due to
a square root singularity the Ward identities are more
complicated than in the NS sector. They read:
\begin{eqnarray}
\label{wardid1}
&&\hspace{-51pt}
\sum_{p=0}^{\infty}
\Big(\!
\begin{array}{c}
\scriptstyle n+\frac12\\[-8pt] \scriptstyle p
\end{array}
\!\Big)\,
z^{n+\frac12 -p}\
\langle \phi(\infty,\infty) (S_{p} R^{\epsilon})(z,\bar z)R^{\epsilon'}(0,0) \rangle
\\
= &&
\nonumber
\sum_{p=0}^{\infty}
\Big(\!
\begin{array}{c}
\scriptstyle \frac12\\[-8pt] \scriptstyle p
\end{array}
\!\Big)\,
(-z)^p \
 \langle (S_{p-n - \frac12} \phi)(\infty,\infty) R^{\epsilon}(z,\bar z) R^{\epsilon'}(0,0) \rangle
 \\
 \nonumber
&&
- i \epsilon  \sum_{p=0}^{\infty}
\Big(\!
\begin{array}{c}
\scriptstyle \frac12\\[-8pt] \scriptstyle p
\end{array}
\!\Big)\,
(-1)^p z^{\frac12 -p}\
\langle \phi(\infty,\infty) R^{\epsilon}(z,\bar z)  (S_{n+p}R^{\epsilon'})(0,0)\rangle,
\\
\label{wardid2}
&& \hspace{-50pt}
\sum_{p=0}^{\infty}
\Big(\!
\begin{array}{c}
\scriptstyle \frac12\\[-8pt] \scriptstyle p
\end{array}
\!\Big)\,
z^{\frac12 -p} \
\langle\phi(\infty,\infty) (S_{p-n} R^{\epsilon})(z,\bar z)  R^{\epsilon'}(0,0) \rangle
\\
\nonumber
= &&
 \sum_{p=0}^{\infty}
\Big(\!
\begin{array}{c}
\scriptstyle \frac12-n\\[-8pt] \scriptstyle p
\end{array}
\!\Big)\,
 (-z)^p \
 \langle ( S_{p+n - \frac12} \phi)(\infty,\infty) R^{\epsilon}(z,\bar z) R^{\epsilon'}(0,0) \rangle
 \\
 \nonumber
&&
- i \epsilon  \sum_{p=0}^{\infty}
\Big(\!
\begin{array}{c}
\scriptstyle \frac12-n\\[-8pt] \scriptstyle p
\end{array}
\!\Big)\,
(-1)^{n+p}  z^{-n+ \frac12 -p}\
\langle \phi(\infty,\infty) R^{\epsilon}(z,\bar z)  (S_{p}R^{\epsilon'})(0,0) \rangle.
\end{eqnarray}
The corresponding relations hold in the anti-holomorphic  sector.

Our aim in this section is to decompose a general 3-point functions onto
left and right 3-point conformal blocks. The first complication arises
due to the block structure of field operators
$$
\Phi = \left[\begin{array}{c|c}
\Phi_{\scriptscriptstyle \rm NN} & 0\\
\hline
0 & \Phi_{\scriptscriptstyle \rm RR}\;
\end{array}\right],
\hskip 1cm
R^\pm = \left[\begin{array}{c|c}
0 & R^\pm_{\scriptscriptstyle \rm NR}\\
\hline
R^\pm_{\scriptscriptstyle \rm RN} & 0
\end{array}\right],
$$
with respect to the direct sum decomposition $ {\cal H}= {\cal
H}_{\scriptscriptstyle \rm NS}\oplus  {\cal
H}_{\scriptscriptstyle\rm R} $ of the space of states.

As in the NS case \cite{Hadasz:2006sb} we define the chiral vertex operators in terms of
3-linear forms. In the R-R sector the form
\begin{eqnarray*}
\varrho{^{\Delta_3 \Delta_2 \Delta_1}_{\scriptscriptstyle RR}}(\,\dots\,;z) &:&
{\cal W}_{\Delta_3}\times {\cal V}_{\Delta_2} \times {\cal W}_{\Delta_1} \ \mapsto \ \mathbb{C}\
\end{eqnarray*}
satisfies the relations\footnote{The chiral Ward identities for the Virasoro generators
$L_n$ are the same in all sectors. The corresponding formulae can be found in
\cite{Hadasz:2006sb}. For the sake of simplicity we suppress all the $L_{-n}$ excitations dependence. }
\begin{eqnarray*}
\varrho{^{\Delta_3 \Delta_2 \Delta_1}_{\scriptscriptstyle RR}}
(S_{-n} \eta_3,\xi_2,\eta_1;z)
&=&
(-1)^{|\eta_1|+|\eta_3|+1}
\varrho{^{\Delta_3 \Delta_2 \Delta_1}_{\scriptscriptstyle RR}}
( \eta_3,\xi_2,S_n\eta_1;z)
\\
\nonumber &+&
\sum\limits_{k=-\frac12}^{\infty}
 \Big(\!
\begin{array}[c]{c}
\scriptstyle n+{1\over 2}\\[-7pt]
\scriptstyle k+\frac12
\end{array}
\!\Big)
  z^{n-k}
 \varrho{^{\Delta_3 \Delta_2 \Delta_1}_{\scriptscriptstyle RR}}
(\eta_3,S_{k}\xi_2,\eta_1;z),
\\
\sum_{p=0}^{\infty}
\Big(\!
\begin{array}{c}
\scriptstyle \frac12\\[-8pt] \scriptstyle p
\end{array}
\!\Big)\,
z^{\frac12 -p} \
\varrho{^{\Delta_3 \Delta_2 \Delta_1}_{\scriptscriptstyle RR}}
(\eta_3,S_{p-k} \xi_2,\eta_1;z )
&=&\\
&& \hspace{-90pt}
 \sum_{p=0}^{\infty}
\Big(\!
\begin{array}{c}
\scriptstyle \frac12-k\\[-8pt] \scriptstyle p
\end{array}
\!\Big)\,
(-z)^p \
\varrho{^{\Delta_3 \Delta_2 \Delta_1}_{\scriptscriptstyle RR}}
(S_{p+k - \frac12} \eta_3, \xi_2,\eta_1;z )
 \\
 &&\hspace{-90pt} - (-1)^{|\eta_3|+|\eta_1|+1 } \sum_{p=0}^{\infty}
\Big(\!
\begin{array}{c}
\scriptstyle \frac12-k\\[-8pt] \scriptstyle p
\end{array}
\!\Big)\,
(-z)^{ \frac12 -k-p}
 \varrho{^{\Delta_3 \Delta_2 \Delta_1}_{\scriptscriptstyle RR}}
(\eta_3,\xi_2,S_{p}\eta_1 ;z),
\end{eqnarray*}
where $|\eta_i|$ denote parities of Ramond states.
The form is determined by the relations above up to  four independent constants.
It is relevant for blocks with R intermediate states which are not considered in the present paper.

 In the NS-R and R-NS sectors the forms (anti-linear in the left argument and linear in the central and the right ones)
\begin{eqnarray*}
\varrho^{\Delta_3 \Delta_2 \Delta_1}_{ \scriptscriptstyle NR}(\,\dots\,;z) &:&
{\cal V}_{\Delta_3}\times {\cal W}_{\Delta_2} \times {\cal W}_{\Delta_1} \ \mapsto \ \mathbb{C},
\\
\varrho{^{\Delta_3 \Delta_2 \Delta_1}_{\scriptscriptstyle RN}}(\,\dots\,;z)
&:&
{\cal W}_{\Delta_3}\times {\cal W}_{\Delta_2} \times {\cal V}_{\Delta_1} \ \mapsto \ \mathbb{C},
\end{eqnarray*}
are defined by
\begin{eqnarray}
\nonumber
&&\hspace{-50pt} \sum_{p=0}^{\infty}
\Big(\!
\begin{array}{c}
\scriptstyle n+\frac12\\[-8pt] \scriptstyle p
\end{array}
\!\Big)\,
z^{n+\frac12 -p} \
\varrho{^{\Delta_3 \Delta_2 \Delta_1}_{\scriptscriptstyle NR}}
(\xi_3,S_{p} \eta_2,\eta_1 ;z)
\\ =&&
 \nonumber
\sum_{p=0}^{\infty}
\Big(\!
\begin{array}{c}
\scriptstyle \frac12\\[-8pt] \scriptstyle p
\end{array}
\!\Big)\,
(-z)^p \
\varrho{^{\Delta_3 \Delta_2 \Delta_1}_{\scriptscriptstyle NR}}
(S_{p-n - \frac12} \xi_3, \eta_2,\eta_1 ;z)
 \\
 \nonumber
&-& i (-1)^{|\xi_3|+|\eta_1|+1}  \sum_{p=0}^{\infty}
\Big(\!
\begin{array}{c}
\scriptstyle \frac12\\[-8pt] \scriptstyle p
\end{array}
\!\Big)\,
 (-1)^p \ z^{\frac12 -p}
\varrho{^{\Delta_3 \Delta_2 \Delta_1}_{\scriptscriptstyle NR}}
(\xi_3, \eta_2,S_{n+p}\eta_1;z )
 ,
\\
 \label{Ward}
&& \hspace{-50pt} \sum_{p=0}^{\infty}
\Big(\!
\begin{array}{c}
\scriptstyle \frac12\\[-8pt] \scriptstyle p
\end{array}
\!\Big)\,
z^{\frac12 -p} \
\varrho{^{\Delta_3 \Delta_2 \Delta_1}_{\scriptscriptstyle NR}}
(\xi_3,S_{p-n} \eta_2,\eta_1 ;z)
\\
 \nonumber
=&&
 \sum_{p=0}^{\infty}
\Big(\!
\begin{array}{c}
\scriptstyle \frac12-n\\[-8pt] \scriptstyle p
\end{array}
\!\Big)\,
 (-z)^p \
 \varrho{^{\Delta_3 \Delta_2 \Delta_1}_{\scriptscriptstyle NR}}
(S_{p+n - \frac12} \xi_3, \eta_2,\eta_1 ;z)
 \\
  \nonumber
 &-&i (-1)^{|\xi_3|+|\eta_1|+1 } \sum_{p=0}^{\infty}
\Big(\!
\begin{array}{c}
\scriptstyle \frac12-n\\[-8pt] \scriptstyle p
\end{array}
\!\Big)\,
 (-1)^{n+p}  z^{ \frac12-n -p}
 \varrho{^{\Delta_3 \Delta_2 \Delta_1}_{\scriptscriptstyle NR}}
(\xi_3,\eta_2,S_{p}\eta_1 ;z)
\end{eqnarray}
and
\begin{eqnarray}\label{Ward_rhoRN}
 \nonumber
&&\hspace{-30pt} \sum_{p=0}^{\infty}
\Big(\!
\begin{array}{c}
\scriptstyle -n\\[-8pt] \scriptstyle p
\end{array}
\!\Big)\,
z^{-p-n} \
\varrho^{\Delta_3 \Delta_2 \Delta_1}_{\scriptscriptstyle \rm RN}
(\eta_3,S_{p} \eta_2,\xi_1;z )
=\sum_{p=0}^{\infty}
\Big(\!
\begin{array}{c}
\scriptstyle \frac12\\[-8pt] \scriptstyle p
\end{array}
\!\Big)\,
(-z)^p \
\varrho^{\Delta_3 \Delta_2 \Delta_1}_{\scriptscriptstyle \rm RN}
(S_{n+p} \eta_3, \eta_2,\xi_1;z )
 \\
&& - i (-1)^{|\eta_3|+|\xi_1|+1}  \sum_{p=0}^{\infty}
\Big(\!
\begin{array}{c}
\scriptstyle \frac12\\[-8pt] \scriptstyle p
\end{array}
\!\Big)\,
(-1)^p \ z^{\frac12 -p}
\varrho^{\Delta_3 \Delta_2 \Delta_1}_{\scriptscriptstyle \rm RN}
(\eta_3, \eta_2,S_{p-n-{1\over 2}}\xi_1 ;z) ,
\\  \nonumber
&& \hspace{-30pt}
\varrho^{\Delta_3 \Delta_2 \Delta_1}_{\scriptscriptstyle \rm RN}
(\eta_3,S_{-n} \eta_2,\xi_1;z)
= \sum_{p=0}^{\infty}
\Big(\!
\begin{array}{c}
\scriptstyle \frac12-n\\[-8pt] \scriptstyle p
\end{array}
\!\Big)\,
(-z)^p \
\varrho^{\Delta_3 \Delta_2 \Delta_1}_{\scriptscriptstyle \rm RN}
(S_{p+n } \eta_3, \eta_2,\xi_1;z)
 \\  \nonumber
 && -i (-1)^{|\eta_3|+|\xi_1|+1 } \sum_{p=0}^{\infty}
\Big(\!
\begin{array}{c}
\scriptstyle \frac12-n\\[-8pt] \scriptstyle p
\end{array}
\!\Big)\,
 (-1)^{n+p}\,z^{ \frac12 -n-p}
 \varrho^{\Delta_3 \Delta_2 \Delta_1}_{\scriptscriptstyle \rm RN}
(\eta_3,\eta_2,S_{p-\frac12 }\xi_1;z ),
\end{eqnarray}
respectively.
They are
almost completely determined by these relations. In particular, for
$L_0$-eingenstates,  $ L_0\,\xi_i  = \Delta_i(\xi_i)\xi_i,\;i=1,3$,
$ L_0\,\eta_j  = \Delta_j(\eta_j)\,\eta_j,\;j=1,2,3$ one has:
\begin{equation}
\label{z:dep}
\begin{array}{rcl}
\varrho{^{\Delta_3 \Delta_2 \Delta_1}_{\scriptscriptstyle NR}
(\xi_3,\eta_2,\eta_1;z)}
&=& z^{\Delta_3(\xi_3)- \Delta_2(\eta_2)-\Delta_1(\eta_1)}\
\varrho{^{\Delta_3 \Delta_2 \Delta_1}_{\scriptscriptstyle NR}}
(\xi_3,\eta_2,\eta_1;1),
\\
 \varrho{^{\Delta_3 \Delta_2 \Delta_1}_{\scriptscriptstyle RN}}
(\eta_3,\eta_2,\xi_1;z)
 &=&
 z^{\Delta_3(\eta_3)- \Delta_2(\eta_2)-\Delta_1(\xi_1)}\
\varrho{^{\Delta_3 \Delta_2 \Delta_1}_{\scriptscriptstyle RN}}
(\eta_3,\eta_2,\xi_1;1).
\end{array}
\end{equation}
 As in the R-R sector the forms depend on
four rather than two arbitrary constants. We define the forms $
\rho^{ij}_{\scriptscriptstyle NR}, \rho^{ij}_{\scriptscriptstyle RN}\,; i,j =\pm$ as coefficients
in front of these constants:
\begin{eqnarray}
\nonumber
&&
\begin{array}{rcl}
\varrho{^{\Delta_3 \Delta_2 \Delta_1}_{\scriptscriptstyle NR}}(\xi_3,\eta_2,\eta_1;z)
&=&
 \rho^{++}_{\scriptscriptstyle NR}(\xi_3,\eta_2,\eta_1;z)
 \varrho{^{\Delta_3 \Delta_2 \Delta_1}_{\scriptscriptstyle NR}}(\nu_3, w^+_2, w^+_1;1)\nonumber \\
&+&\rho^{+-}_{\scriptscriptstyle NR}(\xi_3,\eta_2,\eta_1;z)
 \varrho{^{\Delta_3 \Delta_2 \Delta_1}_{\scriptscriptstyle NR}}(\nu_3, w^+_2, w^-_1;1)\\
&+& \rho^{-+}_{\scriptscriptstyle NR}(\xi_3,\eta_2,\eta_1;z)
 \varrho{^{\Delta_3 \Delta_2 \Delta_1}_{\scriptscriptstyle NR}}(\nu_3, w^-_2, w^+_1;1)\nonumber\\
&+&\rho^{--}_{\scriptscriptstyle NR}(\xi_3,\eta_2,\eta_1;z)
 \varrho{^{\Delta_3 \Delta_2 \Delta_1}_{\scriptscriptstyle NR}}(\nu_3, w^-_2, w^-_1;1)\ ,\nonumber
\end{array}
\\[-6pt]
&&
\label{def_blocks3}
\\[-6pt]
\nonumber
&&
\begin{array}{rcl}
\varrho^{\Delta_3 \Delta_2 \Delta_1}_{\scriptscriptstyle  RN}(\eta_3,\eta_2,\xi_1;z)
&=& 
 \rho^{++}_{\scriptscriptstyle  RN}(\eta_3,\eta_2,\xi_1;z)
 \varrho_{\scriptscriptstyle RN}^{\Delta_3 \Delta_2 \Delta_1}(w^+_3, w^+_2, \nu_1;1)\\
&+& \rho^{+-}_{\scriptscriptstyle  RN}(\eta_3,\eta_2,\xi_1;z)
 \varrho_{\scriptscriptstyle  RN}^{\Delta_3 \Delta_2 \Delta_1}(w^+_3, w^-_2, \nu_1;1 )
 \\
&+& \rho^{-+}_{\scriptscriptstyle \rm RN}(\eta_3,\eta_2,\xi_1;z)
 \varrho_{\scriptscriptstyle  RN}^{\Delta_3 \Delta_2 \Delta_1}(w^-_3, w^+_2, \nu_1;1)\\
&+& \rho^{--}_{\scriptscriptstyle  RN}(\eta_3,\eta_2,\xi_1;z)
\varrho_{\scriptscriptstyle  RN}^{\Delta_3 \Delta_2 \Delta_1}(w^-_3, w^-_2, \nu_1;1).
\end{array}
\end{eqnarray}
From (\ref{z:dep}) one easily gets
\begin{eqnarray*}
\rho^{ij}_{\scriptscriptstyle NR}(\xi_3,\eta_2,\eta_1;z)&=&
z^{\Delta_3(\xi_3)- \Delta_2(\eta_2)-\Delta_1(\eta_1)}
\rho^{ij}_{\scriptscriptstyle NR}(\xi_3,\eta_2,\eta_1),
\\
\rho^{ij}_{\scriptscriptstyle RN}(\eta_3,\eta_2,\xi_1;z)
&=&
z^{\Delta_3(\eta_3)- \Delta_2(\eta_2)-\Delta_1(\xi_1)}
\rho^{ij}_{\scriptscriptstyle RN}(\eta_3,\eta_2,\xi_1).
\end{eqnarray*}
where $\rho^{ij}_{\scriptscriptstyle NR}(\xi_3,\eta_2,\eta_1)=\rho^{ij}_{\scriptscriptstyle NR}(\xi_3,\eta_2,\eta_1;1)$
and a similar notation in the RN sector is assumed.
Analyzing the Ward identities (\ref{Ward}), (\ref{Ward_rhoRN}) one can derive the relations:
\begin{equation}
\label{tNR1}
\begin{array}{rcl}
\rho^{+-}_{\scriptscriptstyle NR }(S_{-I}\nu, w_2^-, S_{-J}w_1^+) &=&
\!i\, \rho^{--}_{\scriptscriptstyle NR}(S_{-I}\nu, w_2^+, S_{-J}w_1^+),
\\
\rho^{-+}_{\scriptscriptstyle NR}(S_{-I}\nu, w_2^-, S_{-J} w_1^+) &=&
\ \rho^{++}_{\scriptscriptstyle NR}(S_{-I}\nu, w_2^+, S_{-J}w_1^+),
\\
\rho^{++}_{\scriptscriptstyle NR}(S_{-I}\nu, w_2^-, S_{-J}w_1^+)
&=& \!i\, \rho^{-+}_{\scriptscriptstyle NR}(S_{-I}\nu, w_2^+, S_{-J}w_1^+),
\\
\rho^{--}_{\scriptscriptstyle NR}(S_{-I}\nu, w_2^-, S_{-J} w_1^+) &=&
\ \rho^{+-}_{\scriptscriptstyle NR}(S_{-I}\nu, w_2^+, S_{-J}w_1^+),
\end{array}
\end{equation}
\begin{equation}
\label{tRN1}
\begin{array}{rcr}
\rho^{+-}_{\scriptscriptstyle  RN  }(S_{J}w_3^+, w_2^-, S_{I}\nu) &=&
 \rho^{++}_{\scriptscriptstyle  RN }( S_{J}w_3^+, w_2^+,S_{I}\nu)  ,
\\
\rho^{-+}_{\scriptscriptstyle  RN  }(S_{J}w_3^+, w_2^-, S_{I}\nu) &=&
\!i\, \rho^{--}_{\scriptscriptstyle  RN }( S_{J}w_3^+, w_2^+,S_{I}\nu) ,
\\
\rho^{++}_{\scriptscriptstyle  RN  }(S_{J}w_3^+, w_2^-, S_{I}\nu) &=&
\!i\,\rho^{+-}_{\scriptscriptstyle  RN }( S_{J}w_3^+, w_2^+,S_{I}\nu)   ,
\\
\rho^{--}_{\scriptscriptstyle  RN  }(S_{J}w_3^+, w_2^-, S_{I}\nu) &=&
 \rho^{-+}_{\scriptscriptstyle  RN }( S_{J}w_3^+, w_2^+,S_{I}\nu)  ,
\end{array}
\end{equation}
and
\begin{equation}
\label{tNR2}
\begin{array}{rcr}
\rho^{+-}_{\scriptscriptstyle \rm NR }(S_{I}\nu, w_2^+, S_{J}w_1^-) &=&
\;\;(-1)^{\# J}\, \rho^{++}_{\scriptscriptstyle \rm NR }(S_{I}\nu, w_2^+, S_{J}w_1^+),
\\
\rho^{-+}_{\scriptscriptstyle \rm NR }(S_{I}\nu, w_2^+, S_{J} w_1^-) &=&
- i\, (-1)^{\# J}\, \rho^{--}_{\scriptscriptstyle \rm NR }(S_{I}\nu, w_2^+, S_{J}w_1^+),
\\
\rho^{++}_{\scriptscriptstyle \rm NR }(S_{I}\nu, w_2^+, S_{J}w_1^-) &=&
- i\, (-1)^{\# J}\, \rho^{+-}_{\scriptscriptstyle \rm NR }(S_{I}\nu, w_2^+, S_{J}w_1^+),
\\
\rho^{--}_{\scriptscriptstyle \rm NR }(S_{I}\nu, w_2^+, S_{J} w_1^-) &=&
\;\;(-1)^{\# J}\, \rho^{-+}_{\scriptscriptstyle \rm NR }(S_{I}\nu, w_2^+, S_{J}w_1^+),
\end{array}
\end{equation}
\begin{equation}
\label{tRN2}
\begin{array}{rcr}
\rho^{+-}_{\scriptscriptstyle \rm RN  }(S_{J}w_3^-, w_2^+, S_{I}\nu) &=&
- i\,(-1)^{\# I}\,
\rho^{--}_{\scriptscriptstyle \rm RN }( S_{J}w_3^+, w_2^+,S_{I}\nu)    ,
\\
\rho^{-+}_{\scriptscriptstyle \rm RN  }(S_{J}w_3^-, w_2^+, S_{I}\nu) &=&
(-1)^{\# I}\,
\rho^{++}_{\scriptscriptstyle \rm RN }( S_{J}w_3^+, w_2^+,S_{I}\nu)  ,
\\
\rho^{++}_{\scriptscriptstyle \rm RN  }(S_{J}w_3^-, w_2^+, S_{I}\nu) &=&
-i\,(-1)^{\# I}\,
\rho^{-+}_{\scriptscriptstyle \rm RN }( S_{J}w_3^+, w_2^+,S_{I}\nu) ,
\\
\rho^{--}_{\scriptscriptstyle \rm RN  }(S_{J}w_3^-, w_2^+, S_{I}\nu) &=&
(-1)^{\# I}\,
 \rho^{+-}_{\scriptscriptstyle \rm RN }( S_{J}w_3^+, w_2^+,S_{I}\nu)  ,  \\[5pt]
\end{array}
\end{equation}
where $\# I$ denotes the number of indices in the multi-index $I$.

We shall now briefly analyze how the global parity
requirements along with the ``small'' representation
reduces the number of independent constants in the matrix elements of Ramond fields
from eight to two.
We shall start with non normalized chiral vertex operators
$V^\pm_{\rm{\scriptscriptstyle NR}\rm e},V^\pm_{\rm{\scriptscriptstyle NR}\rm o}$
 in the NS-R sector. They are
defined in terms of their matrix elements by the form $ \varrho{^{\Delta_3 \Delta_2 \Delta_1}_{\scriptscriptstyle NR}}$
$$
\begin{array}{ccccccl}
\langle \xi_3|V^\pm_{\scriptscriptstyle \rm NR e}(z)|\eta_1\rangle &=&
\varrho{^{\Delta_3 \Delta_2 \Delta_1}_{\scriptscriptstyle  NR}}(\xi_3,w^\pm,\eta_1;z),
& ~~~ & |\xi_3|+ |\eta_1| &\in& 2{\mathbb N},\\
\langle \xi_3|V^\pm_{\rm {\scriptscriptstyle NR o}}(z)|\eta_1\rangle&=&
\varrho{^{\Delta_3 \Delta_2 \Delta_1}_{\scriptscriptstyle NR}}(\xi_3,w^\pm,\eta_1;z),
 &  & |\xi_3|+ |\eta_1| &\in& 2{\mathbb N}+1. \\
\end{array}
$$
From the construction of the highest-weight vectors $w^\pm_{\Delta,\bar \Delta}$ (\ref{basis0})
one may expect the following form of the Ramond fields
\begin{eqnarray*}
R^+_{\scriptscriptstyle \rm NR}&=&
A V^+_{\scriptscriptstyle \rm NR e} \otimes \bar V^+_{\scriptscriptstyle \rm NR e}+
B V^+_{\scriptscriptstyle \rm NR o} \otimes \bar V^+_{\scriptscriptstyle \rm NR o}+
i B V^-_{\scriptscriptstyle \rm NR e} \otimes \bar V^-_{\scriptscriptstyle \rm NR e}
-i A V^-_{\scriptscriptstyle \rm NR o} \otimes \bar V^-_{\scriptscriptstyle \rm NR o}\,,
\\
R^-_{\scriptscriptstyle \rm NR}&=&
A V^+_{\scriptscriptstyle \rm NR e} \otimes \bar V^-_{\scriptscriptstyle \rm NR o}
- B V^+_{\scriptscriptstyle \rm NR o} \otimes \bar V^-_{\scriptscriptstyle \rm NR e}+
B V^-_{\scriptscriptstyle \rm NR e} \otimes \bar V^+_{\scriptscriptstyle \rm NR o}+
A V^-_{\scriptscriptstyle \rm NR o} \otimes \bar V^+_{\scriptscriptstyle \rm NR e}\,,
\end{eqnarray*}
where the coefficients are fixed up to $A$ and $B$ by relations (\ref{strcon}).
The independent structure constants $C^\pm$ (\ref{constant1}) are expressed in terms of
$A,B$ and constants hidden in the forms $\varrho{^{\Delta_3 \Delta_2 \Delta_1}_{\scriptscriptstyle NR}}$,
$\bar \varrho{^{\bar \Delta_3 \bar \Delta_2 \bar \Delta_1}_{\scriptscriptstyle NR}}$ as follows:
\begin{eqnarray}
\nonumber
C^+ &=&  A\, \varrho_{\scriptscriptstyle NR}(\nu, w^+, w^+;1)\, \bar\varrho_{\scriptscriptstyle NR}(\bar\nu, \bar w^+, \bar w^+;1)
         + i B\, \varrho_{\scriptscriptstyle NR}(\nu,  w^-, w^+;1)\, \bar\varrho_{\scriptscriptstyle NR}(\bar \nu, \bar  w^-, \bar w^+;1)
\\ [4pt] \nonumber
&+&
        iB\,  \varrho_{\scriptscriptstyle NR}(\nu, w^+,  w^-;1)\, \bar\varrho_{\scriptscriptstyle NR}(\bar \nu, \bar w^+,\bar w^-;1)
        +A\, \varrho_{\scriptscriptstyle NR}(\nu, w^-, w^-;1) \,
        \bar\varrho_{\scriptscriptstyle NR}(\bar \nu, \bar  w^-,\bar  w^-;1),
\\[-6pt]  \label{C2}\\[-6pt]
\nonumber
C^- &=&\nonumber
 A \, \varrho_{\scriptscriptstyle NR}(\nu, w^+, w^+;1)\, \bar\varrho_{\scriptscriptstyle NR}(\bar \nu,  \bar w^-, \bar  w^-;1)
                +  B \, \varrho_{\scriptscriptstyle NR}(\nu, w^-, w^+)\, \bar\varrho_{\scriptscriptstyle NR}(\bar \nu,\bar  w^+, \bar w^-;1)
                \\[4pt]
&-&  B \, \varrho_{\scriptscriptstyle NR}(\nu,  w^+, w^-;1) \, \bar\varrho_{\scriptscriptstyle NR}(\bar \nu,  \bar w^-, \bar  w^+;1)
            + A \, \varrho_{\scriptscriptstyle NR}(\nu,w^-, w^-;1)\, \bar\varrho_{\scriptscriptstyle NR}(\bar \nu,\bar  w^+,\bar  w^+;1),
            \nonumber
\end{eqnarray}
where $\bar\nu \equiv \nu_{\bar\Delta}$ and $\bar w^{\pm} \equiv w^{\pm}_{\bar\Delta}.$

One can check using relations (\ref{tNR1}), (\ref{tNR2}) that all matrix elements of the
Ramond field $R^\pm_{\scriptscriptstyle NR}$ depend on the arbitrary constants only via the combinations
above. Indeed, using  relations (\ref{tNR2})
one obtains
\begin{eqnarray*}
&& \hskip -2.50cm
\bra{S_{-I} \bar S_{-\bar I} \,\nu_3\otimes\bar \nu _3}  R_2^+  \ket{S_{-J} \bar S_{-\bar J}\, w_{\Delta_1,\bar \Delta_1}^+}\\[4pt]
 &=&  C^{(+)}
\rho^{(+)}_{\scriptscriptstyle \rm  NR e}(S_{-I}\,\nu_3, w_2^+, S_{-J}\,w_1^+)\,
\bar \rho^{(+)}_{\scriptscriptstyle \rm  NR e}(\bar S_{-\bar I}\,\bar\nu_3, \bar w_2^+,\bar S_{-\bar J}\,\bar w_1^+)
 \\ [4pt]
&+&\,C^{(-)}
\rho^{(-)}_{\scriptscriptstyle \rm  NR e}(S_{-I}\, \nu_3, w_2^+, S_{-J}\,w_1^+)\,
\bar \rho^{(-)}_{\scriptscriptstyle \rm  NR e}(\bar S_{-\bar I}\,\bar\nu_3, \bar w_2^+,\bar S_{-\bar J}\,\bar w_1^+)
\\
&& \hskip -2.5cm
\bra{S_{-I} \bar S_{-\bar I} \,\nu_3\otimes\bar \nu _3 }  R_2^-  \ket{S_{-J} \bar S_{-\bar J}\, w_{\Delta_1,\bar \Delta_1}^+}\\[4pt]
 &=&  {(-1)^{|J|}}  C^{(+)}
\rho^{(+)}_{\scriptscriptstyle \rm  NR e}(S_{-I}\,\nu_3, w_2^+, S_{-J}\,w_1^+)\,
\bar \rho^{(+)}_{\scriptscriptstyle \rm  NR o}(\bar S_{-\bar I}\,\bar\nu_3, \bar w_2^+,\bar S_{-\bar J}\,\bar w_1^+)
 \\ [4pt]
&-& {(-1)^{|J|}} C^{(-)}
\rho^{(-)}_{\scriptscriptstyle \rm  NR e}(S_{-I}\, \nu_3, w_2^+, S_{-J}\,w_1^+)\,
\bar \rho^{(-)}_{\scriptscriptstyle \rm  NR o}(\bar S_{-\bar I}\,\bar\nu_3,\bar w_2^+,\bar S_{-\bar J}\,\bar w_1^+)
\end{eqnarray*}
for $2|I|=|J|$ and
\begin{eqnarray*}
&& \hskip -2.0cm   \bra{S_{-I} \bar S_{-\bar I}\, \nu_3\otimes\bar \nu _3 }  R_2^+  \ket{S_{-J} \bar S_{-\bar J}\, w_{\Delta_1,\bar \Delta_1}^+}\\
  & =&  -i {(-1)^{|J|}} C^{(+)}
       \rho^{(+)}_{\scriptscriptstyle \rm NRo}(S_{-I}\, \nu_3, w_2^+, S_{-J}\, w_1^+)\,
       \bar \rho^{(+)}_{\scriptscriptstyle \rm NRo}(\bar S_{-\bar I}\bar\nu_3, \bar w_2^+,\bar  S_{-\bar J}\, \bar w_1^+)
 \\ [4pt]
&&- i \, {(-1)^{|J|} }  C^{(-)}
        \rho^{(-)}_{\scriptscriptstyle \rm NRo}(S_{-I}\,\nu_3, w_2^+, S_{-J}\,w_1^+)\,
       \bar \rho^{(-)}_{\scriptscriptstyle \rm NRo}(\bar  S_{-\bar I}\,\bar\nu_3, \bar w_2^+,\bar  S_{-\bar J}\,\bar w_1^+),
\\[6pt]
&& \hskip -2.0cm   \bra{S_{-I} \bar S_{-\bar I}\, \nu_3\otimes\bar \nu _3 }  R_2^-  \ket{S_{-J} \bar S_{-\bar J}\, w_{\Delta_1,\bar \Delta_1}^+}\\
  & =&   C^{(+)}
       \rho^{(+)}_{\scriptscriptstyle \rm NRo}(S_{-I}\, \nu_3, w_2^+, S_{-J}\, w_1^+)\,
       \bar \rho^{(+)}_{\scriptscriptstyle \rm NRe}(\bar S_{-\bar I}\bar\nu_3, \bar w_2^+,\bar  S_{-\bar J}\, \bar w_1^+)
 \\ [4pt]
&-& C^{(-)}
        \rho^{(-)}_{\scriptscriptstyle \rm NRo}(S_{-I}\,\nu_3, w_2^+, S_{-J}\,w_1^+)\,
       \bar \rho^{(-)}_{\scriptscriptstyle \rm NRe}(\bar  S_{-\bar I}\,\bar\nu_3, \bar w_2^+,\bar  S_{-\bar J}\,\bar w_1^+)
\end{eqnarray*}
for $2|I|=|J|+1$,
where
\begin{eqnarray*}
C^{(\pm)}&=&{ C^+ \pm C^-\over 2},
\\
\rho^{(\pm)}_{\scriptscriptstyle \rm NRe}
&=&
\rho^{++}_{\scriptscriptstyle NR} \pm
        \rho^{--}_{\scriptscriptstyle NR},
\hskip 1cm
\bar \rho^{(\pm)}_{\scriptscriptstyle \rm NRe}
\;=\;
\bar \rho^{++}_{\scriptscriptstyle NR} \pm
        \bar \rho^{--}_{\scriptscriptstyle NR},
\\
\rho^{(\pm)}_{\scriptscriptstyle \rm NRo}
&=&
\rho^{+-}_{\scriptscriptstyle NR} \pm i
        \rho^{-+}_{\scriptscriptstyle NR},
\hskip 1cm
\bar \rho^{(\pm)}_{\scriptscriptstyle \rm NRo}
\;=\;
\bar \rho^{+-}_{\scriptscriptstyle NR} \mp i
        \bar \rho^{-+}_{\scriptscriptstyle NR}.
\end{eqnarray*}
Introducing chiral vertex operators
\begin{equation}
\label{defVNR}
\begin{array}{ccccclllllll}
\langle \xi_3|V^{(\pm)}_{\scriptscriptstyle \rm NR e}(z)|\eta_1\rangle &=&
\rho^{(\pm)}_{\scriptscriptstyle \rm  NR e}(\xi_3,w^+,\eta_1;z),
& ~~~~& |\xi_3|+ |\eta_1| &\in& 2{\mathbb N},\\
\langle \xi_3|V^{(\pm) }_{\rm \scriptscriptstyle NR o}(z)|\eta_1\rangle &=&
\rho{^{(\pm)}_{\scriptscriptstyle\rm NRo}}(\xi_3,w^+,\eta_1;z),
 && |\xi_3|+ |\eta_1| &\in& 2{\mathbb N}+1,
\end{array}
\end{equation}
one thus gets

\begin{eqnarray}
\label{chiraldec}
\nonumber
  R_{\scriptscriptstyle \rm NR}^+
  &=&
{C^{(+)} \over \sqrt2}
  \left( V_{\scriptscriptstyle \rm NRe}^{(+)} \otimes
  \bar V_{\scriptscriptstyle \rm NRe}^{(+)}
            \,-\, i \,
V_{\scriptscriptstyle \rm NRo}^{(+)} \otimes
\bar V_{\scriptscriptstyle \rm NRo}^{(+)} \right)
+
{C^{(-)} \over \sqrt2}
\left( V_{\scriptscriptstyle \rm NRe}^{(-)} \otimes
\bar V_{\scriptscriptstyle \rm NRe}^{(-)}
            \,-\, i \,
V_{\scriptscriptstyle \rm NRo}^{(-)} \otimes
\bar V_{\scriptscriptstyle \rm NRo}^{(-)}\right),
\\[-2pt]
\\[-8pt]
\nonumber
R_{\scriptscriptstyle \rm NR}^-
&=&
{C^{(+)} \over \sqrt2}
\left( V_{\scriptscriptstyle \rm NRe}^{(+)} \otimes
\bar V_{\scriptscriptstyle \rm NRo}^{(+)}
            \,+\,
V_{\scriptscriptstyle \rm NRo}^{(+)} \otimes
\bar V_{\scriptscriptstyle \rm NRe}^{(+)} \right)
-
{C^{(-)} \over \sqrt2}
\left( V_{\scriptscriptstyle \rm NRe}^{(-)} \otimes
\bar V_{\scriptscriptstyle \rm NRo}^{(-)}
            \,+\,
V_{\scriptscriptstyle \rm NRo}^{(-)} \otimes
\bar V_{\scriptscriptstyle \rm NRe}^{(-)} \right).
\end{eqnarray}
In the R-NS sector a similar analysis  yields the 3-point blocks
\begin{equation}
\label{3ptBRN}
\begin{array}{lllllllllllllll}
\rho^{(\pm)}_{\scriptscriptstyle \rm RN e}
&=&
\rho^{++}_{\scriptscriptstyle \rm RN} \pm
        \rho^{--}_{\scriptscriptstyle \rm RN},
&&\;\;\;\;
\bar \rho^{(\pm)}_{\scriptscriptstyle \rm RN e}
&=&
\bar \rho^{++}_{\scriptscriptstyle \rm RN} \pm
        \bar \rho^{--}_{\scriptscriptstyle \rm RN},
\\
\rho^{(\pm)}_{\scriptscriptstyle \rm RN o}
&=&
 \rho^{-+}_{\scriptscriptstyle \rm RN} \pm i
        \rho^{+-}_{\scriptscriptstyle \rm RN},
&&\;\;\;\;
\bar \rho^{(\pm)}_{\scriptscriptstyle \rm RN o}
&=&
\bar \rho^{-+}_{\scriptscriptstyle \rm RN} \mp i
        \bar \rho^{+-}_{\scriptscriptstyle \rm RN},
\end{array}
\end{equation}
and the chiral vertex operators
$$
\begin{array}{cccccl}
\langle \eta_3|V^{(\pm)}_{\scriptscriptstyle \rm RN p}(z)|\xi_1\rangle &=&
\rho^{(\pm)}_{\scriptscriptstyle \rm  RN p}(\eta_3,w^+,\xi_1;z).
\end{array}
$$
In terms of these operators  the fields $R^\pm_{\scriptscriptstyle \rm RN}$
take the form:
\begin{eqnarray}
\label{R_NRvertex}
\nonumber
  R_{\scriptscriptstyle \rm RN}^+
  &=&
{C^{(+)} \over \sqrt2}
  \left( V_{\scriptscriptstyle \rm RNe}^{(+)} \otimes
  \bar V_{\scriptscriptstyle \rm RNe}^{(+)}
            \,-\, i \,
V_{\scriptscriptstyle \rm RNo}^{(+)} \otimes
\bar V_{\scriptscriptstyle \rm RNo}^{(+)} \right)
+
{C^{(-)} \over \sqrt2}
\left( V_{\scriptscriptstyle \rm RNe}^{(-)} \otimes
\bar V_{\scriptscriptstyle \rm RNe}^{(-)}
            \,-\, i \,
V_{\scriptscriptstyle \rm RNo}^{(-)} \otimes
\bar V_{\scriptscriptstyle \rm RNo}^{(-)}\right),
\\[-2pt]
\\[-8pt]
\nonumber
R_{\scriptscriptstyle \rm RN}^-
&=&
{C^{(+)} \over \sqrt2}
\left( V_{\scriptscriptstyle \rm RNe}^{(+)} \otimes
\bar V_{\scriptscriptstyle \rm RNo}^{(+)}
            \,+\,
V_{\scriptscriptstyle \rm NRo}^{(+)} \otimes
\bar V_{\scriptscriptstyle \rm RNe}^{(+)} \right)
-
{C^{(-)} \over \sqrt2}
\left( V_{\scriptscriptstyle \rm RNe}^{(-)} \otimes
\bar V_{\scriptscriptstyle \rm RNo}^{(-)}
            \,+\,
V_{\scriptscriptstyle \rm RNo}^{(-)} \otimes
\bar V_{\scriptscriptstyle \rm RNe}^{(-)} \right).
\end{eqnarray}
Let us observe that
$$
w^+_\Delta \otimes  w^+_{\bar\Delta} +i\,
 w^-_\Delta \otimes  w^-_{\bar\Delta},
\hskip 10pt
 w^+_\Delta \otimes  w^-_{\bar\Delta} -
 w^-_\Delta \otimes  w^+_{\bar\Delta}\;\in\;{\rm ker} R_{\scriptscriptstyle \rm NR}^\pm\,,
$$
 hence the ``small representation'' is an invariant subspace of the
full Ramond fields $R^\pm$.

The forms
$\rho^{(\pm)}_{\scriptscriptstyle \rm NR f},\rho^{(\pm)}_{\scriptscriptstyle \rm RN f}$
depend on the sign of $\beta_2$ in a very simple way:
\begin{equation}
\label{signdep}
\begin{array}{rcr}
\rho^{(\pm)}_{\scriptscriptstyle \rm NR p}(S_{-I}\nu,w^+_{-\beta_2},S_{-J} w^+_1)
&=&
\rho^{(\mp)}_{\scriptscriptstyle \rm NR p}(S_{-I}\nu,w^+_{\beta_2},S_{-J} w^+_1),
\\
\rho^{(\pm)}_{\scriptscriptstyle \rm RN p}(S_{-I}w^+_3,w^+_{-\beta_2},S_{-J} \nu)
&=&
\rho^{(\mp)}_{\scriptscriptstyle \rm RN p}(S_{-I}w^+_3,w^+_{\beta_2},S_{-J} \nu).
\end{array}
\end{equation}
One can assume that $C^+$ does not depend on the signs of $\beta$-s involved. Then  chiral decompositions
(\ref{chiraldec}), (\ref{R_NRvertex})
imply
\begin{eqnarray}
\label{formula}
R^\epsilon_{-\beta}=\epsilon R^\epsilon_{\beta}\,.
\end{eqnarray}

As a side remark let us mention that decompositions (\ref{chiraldec}), (\ref{R_NRvertex}) can
be easily extended to excited Ramond fields using central arguments
of the forms $\rho^{(\pm)}_{\scriptscriptstyle \rm NR e}, \rho^{(\pm)}_{\scriptscriptstyle \rm NR o}$:
\begin{equation}
\label{defVNRext}
\begin{array}{ccccl}
\begin{array}{rcl}
\langle \xi_3|V^{(\pm)}_{\scriptscriptstyle \rm NR e}(\eta_2)|\eta_1\rangle
&=&
\rho^{(\pm)}_{\scriptscriptstyle \rm  NR e}(\xi_3,\eta_2,\eta_1),
\\
\langle \eta_1|V^{(\pm)}_{\scriptscriptstyle \rm RN e}(\eta_2)|\xi_3\rangle
&=&
\rho^{(\pm)}_{\scriptscriptstyle \rm  RN e}(\eta_1,\eta_2,\xi_3),
\end{array}
& ~~~ & |\xi_3|+ |\eta_1| &\in& 2{\mathbb N},\\[8pt]
\begin{array}{rcl}
\langle \xi_3|V^{(\pm) }_{\rm \scriptscriptstyle NR o}(\eta_2)|\eta_1\rangle &=&
\rho{^{(\pm)}_{\scriptscriptstyle\rm NR o}}(\xi_3,\eta_2,\eta_1),
\\
  \langle \eta_1|V^{(\pm) }_{\rm \scriptscriptstyle RN o}
(\eta_2)|\xi_3\rangle
&=&
\rho{^{(\pm)}_{\scriptscriptstyle\rm RN o}}(\eta_1,\eta_2,\xi_3),
\end{array}
 &  & |\xi_3|+ |\eta_1| &\in& 2{\mathbb N}+1.
\end{array}
\end{equation}
Taking into account the graded tensor product structure:
\begin{eqnarray*}
\langle \xi_3\otimes \bar \xi_3|
V^{(\pm)}_{\scriptscriptstyle \rm NR p}(\eta_2)
\otimes
\bar V^{(\pm)}_{\scriptscriptstyle \rm NR \bar p}(\bar \eta_2)
|\eta_1\otimes \bar \eta_1\rangle
&=&(-1)^{|p|| \bar \xi_3|+|\bar p||\eta_1|}
\rho{^{(\pm)}_{\scriptscriptstyle\rm NRp}}(\xi_3,\eta_2,\eta_1)
\bar \rho{^{(\pm)}_{\scriptscriptstyle\rm NR\bar p}}(\bar\xi_3,\bar\eta_2,\bar\eta_1),
\\[4pt]
\langle \eta_3\otimes \bar \eta_3|
V^{(\pm)}_{\scriptscriptstyle \rm RN p}(\eta_2)
\otimes
\bar V^{(\pm)}_{\scriptscriptstyle \rm RN \bar p}(\bar \eta_2)
|\xi_1\otimes \bar \xi_1\rangle
&=&
(-1)^{|p||{\bar \eta_3}|+|\bar p||\xi_1|}
\rho{^{(\pm)}_{\scriptscriptstyle\rm RN p}}(\eta_3,\eta_2,\xi_1)
\bar \rho{^{(\pm)}_{\scriptscriptstyle\rm RN \bar p}}(\bar\eta_3,\bar\eta_2,\bar\xi_1),
\end{eqnarray*}
where ${\rm p},\bar {\rm p}= {\rm e, o}$ and $|{\rm e}|=0,|{\rm o}|=1$,
one gets for such extension
\begin{eqnarray*}
-iR^+( w^-\otimes \bar w^-)&=& R^+( w^+\otimes \bar w^+),\\
R^+(w^+\otimes \bar w^-) &=& R^+(w^-\otimes \bar w^+)\;=\;R^-(w^+\otimes \bar w^+).
\end{eqnarray*}

\section{Analytic structure of 4-point conformal blocks}

We shall restrict ourselves to the case of correlation functions of four Ramond fields.
Te structure of the 3-point conformal blocks analyzed in the previous section
suggests the following definition:
\begin{eqnarray*}
&&
F^{f}_{c, \Delta}
\left[^{\pm \beta_3 \; \pm \beta_2}_{\;\;\;\beta_4
\,\;\;\;\beta_1} \right] \; =
\hspace*{-20pt}
\begin{array}[t]{c}
{\displaystyle\sum} \\[2pt]
{\scriptstyle
|K|+|M| = |L|+|N| = f
}
\end{array}
\hspace*{-20pt}
\rho^{(\pm)}_{\scriptscriptstyle {\rm RN} |f|} (w^+_{4} ,w^+_{3} ,\nu_{\Delta,KM})
\ \left[B^{f}_{c, \Delta}\right]^{KM,LN}  \rho^{(\pm)}_{\scriptscriptstyle {\rm NR} |f|}
   (\nu_{\Delta,LN},  w^+_{2} , w^+_{1} ),
\end{eqnarray*}
where ${|f|}={\rm e}$ for $f\in \mathbb{N}$, ${|f|}={\rm o}$ for $f\in \mathbb{N}-{1\over 2}$,
 $\nu_{\Delta,KM}$ is the standard basis in the NS Verma module ${\cal V}_{c,\Delta}$, and
$\left[B^{f}_{c, \Delta}\right]^{KM,LN}$ denotes the inverse to the Gram matrix with respect to this basis
on the level $f\in {1\over 2}\mathbb{N}$. One has four even:
\begin{eqnarray}
\label{evenblock}
\mathcal{F}^1_{\Delta}
\left[^{\pm \beta_3 \; \pm \beta_2}_{\;\;\;\beta_4
\,\;\;\;\beta_1} \right]
(z)
    &=&
z^{\Delta - \Delta_2 - \Delta_1} \left( 1 +
\sum_{m\in \mathbb{N}} z^m
    F^m_{c, \Delta}
    \left[^{\pm \beta_3 \; \pm \beta_2}_{\;\;\;\beta_4
\,\;\;\;\beta_1} \right]
     \right),
\end{eqnarray}
and four odd,
\begin{eqnarray}
\label{oddblock}
\mathcal{F}^{\frac{1}{2}}_{\Delta}
\left[^{\pm \beta_3 \; \pm \beta_2}_{\;\;\;\beta_4
\,\;\;\;\beta_1} \right]
(z)
    &=&
z^{\Delta +{1\over 2}- \Delta_2 - \Delta_1 }
\sum_{k\in \mathbb{N}- \frac{1}{2}} z^{k-{1\over 2}}
        F^k_{c, \Delta}
       \left[^{\pm \beta_3 \; \pm \beta_2}_{\;\;\;\beta_4
\,\;\;\;\beta_1} \right],
\end{eqnarray}
conformal blocks.

It follows from the definition of  blocks' coefficients
$
F^{f}_{c, \Delta}\!
\left[^{\pm \beta_3 \; \pm \beta_2}_{\;\;\;\beta_4
\,\;\;\;\beta_1} \right]
$
 that they are polynomials in $\beta_i$
and rational functions of the intermediate weight $\Delta$ and the central charge $c.$ They
can be expressed  as a sum over the poles in~$\Delta:$
\begin{equation}
\label{first:expansion:Delta}
F^{f}_{c, \Delta}\!
\left[^{\pm \beta_3 \; \pm \beta_2}_{\;\;\;\beta_4
\,\;\;\;\beta_1} \right]
\; = \;
{R}^{f}_{c,\Delta}\!
\left[^{\pm \beta_3 \; \pm \beta_2}_{\;\;\;\beta_4
\,\;\;\;\beta_1} \right]
+
\begin{array}[t]{c}
{\displaystyle\sum} \\[-6pt]
{\scriptscriptstyle
1 < rs \leq 2{f}}
\\[-8pt]
{\scriptscriptstyle
r + s\in 2{\mathbb N}
}
\end{array}
\frac{
{\mathcal R}^{{f}}_{c,\,rs}\!
\left[^{\pm \beta_3 \; \pm \beta_2}_{\;\;\;\beta_4
\,\;\;\;\beta_1} \right]
}
{
\Delta-\Delta_{rs}(c)
}\,,
\end{equation}
 with $\Delta_{rs}(c)$ given by Kac determinant formula for NS Verma modules:
\begin{eqnarray}
\label{delta:rs}
\Delta_{rs}(c)
& = &
-\frac{rs-1}{4} + \frac{1-r^2}{8}b^2 + \frac{1-s^2}{8}\frac{1}{b^2}\,,\hskip 10mm
c=\frac{3}{2} +3\left(b+{1\over b}\right)^2.
\end{eqnarray}

The calculation of the residue at $\Delta_{rs}$ is essentially the same as in
the NS case.
With a suitable choice of basis in ${\cal V}_\Delta$ one gets
\begin{eqnarray}
\label{res:1}
&&
{\mathcal R}^{{f}}_{c,\,rs}\!
\left[^{\pm \beta_3 \; \pm \beta_2}_{\;\;\;\beta_4
\,\;\;\;\beta_1} \right]
 \; = \; A_{rs}(c) \ \times\\
\nonumber
&&
\sum
     \rho^{(\pm)}_{\scriptscriptstyle {\rm RN} |f|}( w^+_4, w^+_3 , S_{-K}L_{-M}\chi_{rs} )
   \ \left[B^{{f}-\frac{rs}{2}}_{c, \Delta_{rs}+\frac{rs}{2}}\right]^{KM,LN} \!
   \rho^{(\pm)}_{\scriptscriptstyle {\rm NR} |f|}
   (S_{-L}L_{-N}\chi_{rs} ,  w^+_2 , w^+_1 ),
\end{eqnarray}
with
\begin{equation}
\label{A:rs:1}
A_{rs}(c)
\; = \;
\lim_{\Delta\to\Delta_{rs}}
\left(\frac{\left\langle\chi_{rs}^\Delta|\chi_{rs}^\Delta\right\rangle}{\Delta - \Delta_{rs}(c)}
\right)^{-1}.
\end{equation}
The coefficients are the same as in the NS-NS sector and their explicit form
has been conjectured in \cite{Belavin:2006} to be:
\begin{equation}
\label{A:rs:2}
A_{rs}(c)
\; = \;
\frac12
\prod_{p=1-r}^r
\prod_{q=1-s}^s
\left(\frac{1}{\sqrt{2}}\left(pb+\frac{q}{b}\right)\right)^{-1}
\hskip -3mm,
\hskip .5cm
p+q \in 2{\mathbb Z}, \; (p,q) \neq (0,0),(r,s).
\end{equation}

The factorization property
of the forms $
\rho^{(\pm)}_{\scriptscriptstyle \rm NR}$ holds in the present case  only on singular vectors.
In the case  ${rs\over 2} \in \mathbb{N}$ one gets
\begin{eqnarray*}
\begin{array}{rcl}
\rho_{\scriptscriptstyle \rm NRe}^{(\pm)}(S_{-I}\chi_{rs}, w_2^+, w_{1}^+)
&=&
\rho_{\scriptscriptstyle \rm NRe}^{(\pm)}(S_{-I}\nu_{\Delta_{rs}+\frac{rs}{2}}, w_2^+, w_{1}^+)
\, \rho_{\scriptscriptstyle \rm NRe}^{(\pm)}(\chi_{rs}, w_2^+, w_{1}^+)
\\      [6pt]
\rho_{\scriptscriptstyle \rm RN e}^{(\pm)}(w_{4}^+, w_3^+,S_{-I}\chi_{rs} )
&=&
\rho_{\scriptscriptstyle \rm RN e}^{(\pm)}(w_{4}^+, w_3^+, S_{-I}\nu_{\Delta_{rs}+\frac{rs}{2}})
\, \rho_{\scriptscriptstyle \rm RN e}^{(\pm)}(w_{4}^+, w_3^+, \chi_{rs})
\end{array}
&,&
|I|\in \mathbb{N},
\\ [6pt]
\begin{array}{rcl}
\rho_{\scriptscriptstyle \rm NRo}^{(\pm)}(S_{-K}\chi_{rs}, w_2^+, w_{1}^+)
&=&
\rho_{\scriptscriptstyle \rm NRo}^{(\pm)}(S_{-K}\nu_{\Delta_{rs}+\frac{rs}{2}}, w_2^+, w_{1}^+)
\,
\rho_{\scriptscriptstyle \rm NRe}^{(\pm)}(\chi_{rs}, w_2^+, w_{1}^+)
\\[6pt]
\rho_{\scriptscriptstyle \rm RNo}^{(\pm)}(w_{4}^+, w_3^+,S_{-K}\chi_{rs})
&=&
\rho_{\scriptscriptstyle \rm RNo}^{(\pm)}(w_{4}^+, w_3^+, S_{-K}\nu_{\Delta_{rs}+\frac{rs}{2}})
\,
\rho_{\scriptscriptstyle \rm RNe}^{(\pm)}(w_{4}^+, w_3^+, \chi_{rs})
\end{array}
&,&
|K|\in \mathbb{N}+{\textstyle {1\over 2}},
\end{eqnarray*}
while in the case ${rs\over 2} \in \mathbb{N} + \frac12$
\begin{eqnarray*}
\begin{array}{rcr}
\rho_{\scriptscriptstyle \rm NRo}^{(\pm)}(S_{-I}\chi_{rs}, w_2^+, w_{1}^+)
&=&
\rho_{\scriptscriptstyle \rm NRe}^{(\mp)}(S_{-I}\nu_{\Delta_{rs}+\frac{rs}{2}}, w_2^+, w_{1}^+)
\,
\rho_{\scriptscriptstyle \rm NRo}^{(\pm)}(\chi_{rs}, w_2^+, w_{1}^+)
\\      [6pt]
\rho_{\scriptscriptstyle \rm RNo}^{(\pm)}(w_{4}^+, w_3^+,S_{-I}\chi_{rs} )
&=&
\rho_{\scriptscriptstyle \rm RNe}^{(\mp)}(w_{4}^+, w_3^+,S_{-I}\nu_{\Delta_{rs}+\frac{rs}{2}} )
\,
\rho_{\scriptscriptstyle \rm RNo}^{(\pm)}( w_{4}^+, w_3^+,\chi_{rs})
\end{array}
&,&
|I|\in \mathbb{N},
\\ [6pt]
\begin{array}{rcr}
\rho_{\scriptscriptstyle \rm NRe}^{(\pm)}(S_{-K}\chi_{rs}, w_2^+, w_{1}^+)
&=&
i     \,
\rho_{\scriptscriptstyle \rm NRo}^{(\mp)}(S_{-K}\nu_{\Delta_{rs}+\frac{rs}{2}}, w_2^+, w_{1}^+)
\,
\rho_{\scriptscriptstyle \rm NRo}^{(\pm)}(\chi_{rs}, w_2^+, w_{1}^+)
\\    [6pt]
\rho_{\scriptscriptstyle \rm RNe}^{(\pm)}(w_{4}^+, w_3^+,S_{-K}\chi_{rs} )
&=&
{-i}   \,
\rho_{\scriptscriptstyle \rm RNo}^{(\mp)}(w_{4}^+, w_3^+,S_{-K}\nu_{\Delta_{rs}+\frac{rs}{2}} )
\,
\rho_{\scriptscriptstyle \rm RNo}^{(\pm)}( w_{4}^+, w_3^+,\chi_{rs})
\end{array}
&,&
|K|\in \mathbb{N}+{\textstyle {1\over 2}}.
\end{eqnarray*}
This yields
\begin{eqnarray}
\label{res:even}
{\mathcal R}^{{f}}_{c,\,rs}
\left[^{\pm \beta_3 \; \pm \beta_2}_{\;\;\;\beta_4
\,\;\;\;\beta_1} \right]
&=&
\\
\nonumber
&&\hspace{-40pt}
A_{rs}(c)\,
\rho_{\scriptscriptstyle \rm RNe}^{(\pm)}(w_4^+ , w_{3}^+,\chi_{rs})
\rho_{\scriptscriptstyle \rm NRe}^{(\pm)}(\chi_{rs}, w_2^+, w_{1}^+)
F^{{f}-\frac{rs}{2}}_{c, \Delta_{rs} + \frac{rs}{2}}
\left[^{\pm \beta_3 \; \pm \beta_2}_{\;\;\;\beta_4
\,\;\;\;\beta_1} \right]
\end{eqnarray}
for $\frac{rs}{2} \in {\mathbb N}\cup \{0\}$,
and
\begin{eqnarray}
\label{res:odd}
{\mathcal R}^{{f}}_{c,\,rs}
\left[^{\pm \beta_3 \; \pm \beta_2}_{\;\;\;\beta_4
\,\;\;\;\beta_1} \right]
&=&
\\
\nonumber
&&\hspace{-40pt}
A_{rs}(c)\,
\rho_{\scriptscriptstyle \rm RNo}^{(\pm)}(w_4^+, w_{3}^+,\chi_{rs} )
\rho_{\scriptscriptstyle \rm NRo}^{(\pm)}(\chi_{rs}, w_2^+, w_{1}^+)
F^{{f}-\frac{rs}{2}}_{c, \Delta_{rs} + \frac{rs}{2}}
\left[^{\mp \beta_3 \; \mp \beta_2}_{\;\;\;\beta_4
\,\;\;\;\beta_1} \right]
\end{eqnarray}
for $\frac{rs}{2} \in {\mathbb N}-\frac12$.

In order to calculate $\rho_{\scriptscriptstyle \rm NRp}^{(\pm)}(\chi_{rs}, w_2^+, w_{1}^+)$,
$\rho_{\scriptscriptstyle \rm RNp}^{(\pm)}(w_4^+, w_{3}^+,\chi_{rs} )$
we shall first consider  the three point correlation functions with degenerate field $\chi_{rs}$ within
the Feigin-Fuchs construction \cite{Bershadsky:1985dq}. In this approach
the Ramond fields are represented by vertex operators in the free superscalar Hilbert space
\begin{equation}
\label{vertexop}
R^+_{\beta,\bar \beta} (z,\bar z)=  {\rm e}^{a \phi(z)+\bar a\bar \phi(\bar z)}\sigma^+(z,\bar z),
\hskip 1cm
R^-_{\beta,\bar \beta} (z,\bar z)=   {\rm e}^{a \phi(z)+\bar a\bar  \phi(\bar z)}\sigma^-(z,\bar z),
\end{equation}
where $a = \frac{Q}{2} - \sqrt{2} \beta $ and $\sigma^\pm$ are the twist operators of the fermionic sector:
\begin{equation}
\label{sigmaope}
\psi(z)\sigma^\pm(z,\bar z) \sim { {\rm e}^{\mp i{\pi\over 4}}\over \sqrt{2(z-w)}} \;\sigma^\mp(z,\bar z).
\end{equation}
The left chiral screening charges are given by:
\[
Q_b = \oint dz \, \psi(z) {\rm e}^{b \phi(z)}, \qquad
Q_{\frac{1}{b}} = \oint dz \, \psi(z) {\rm e}^{\frac{1}{b} \phi(z)},
\]
and the same construction holds in the right sector.
Let us  consider the Feigin-Fuchs representation of
three point functions with various number of left screening charges:
\begin{equation}\label{charges}
\begin{array}{llllll}
C^{\epsilon}_{(\alpha_{rs},\delta), (\beta_2,0), (\beta_1,0)}
&=& \left\langle {\chi_{rs}}R_{\beta_2}^{\epsilon} R_{\beta_1}^{\epsilon} \,
 Q^k_b \, Q^l_{\frac{1}{b}}  \right\rangle,
&
k+l\in 2\mathbb{N}\ ,
& \textstyle \delta=-{1\over 2\sqrt{2}}({1\over b}+b),
\\[6pt]
C^{\epsilon}_{(\alpha_{rs},\delta), (\beta_2,0), (\beta_1,0)}
&=& \left\langle {\chi_{rs}} R_{\beta_2}^{\epsilon} R_{\beta_1}^{\epsilon} \,
 Q^k_b \, Q^l_{\frac{1}{b}} \, \bar Q_b  \right\rangle,\;
&k+l\in2\mathbb{N} +1,\;
& \textstyle \delta={1\over 2\sqrt{2}}({1\over b}-b).
 \end{array}
\end{equation}
The charge conservation implies that the
 structure constants above are non-zero if and only if
the even fusion rules ( $k+l \in 2{\mathbb N} \cup \{0\}$):
 \begin{equation}
\label{efusionRules} \beta_1 + \beta_2 \; =
\; \frac{1}{2 \sqrt{2}} \,(1-r+2k)b + \frac{1}{2 \sqrt{2}} \, (1-s+2l)\frac{1}{b},
\end{equation}
or
the odd fusion rules ( $k+l \in 2{\mathbb N} -1 $):
\begin{equation}
\label{ofusionRules}
\beta_1 + \beta_2 \; =
\; \frac{1}{2 \sqrt{2}} \,(1-r+2k)b + \frac{1}{2 \sqrt{2}} \, (1-s+2l)\frac{1}{b}
\end{equation}
are satisfied ($k,l$ are integers in the range $ 0 \leq k \leq r-1,\ 0 \leq l \leq s-1 $).

In the Feigin-Fuchs representation one can show that for any even integer $n\in 2\mathbb{N}$:
\begin{eqnarray} \label{twist}
\left\langle \psi(w_1) \ldots \psi(w_{n}) \sigma^-(1,1)  \sigma^-(0,0) \right\rangle  \nonumber
&=&-\left\langle \psi(w_1) \ldots \psi(w_{n})
) \sigma^+(1,1) \sigma^+(0,0)\right\rangle,
 \\[4pt]
 \left\langle \psi(w_1) \ldots \psi(w_{n-1}) \, \bar \psi(\bar w) \sigma^-(1,1)
 \sigma^-(0,0) \right\rangle  \nonumber
&=& \left\langle \psi(w_1) \ldots \psi(w_{n-1})\, \bar \psi(\bar w) \sigma^+(1,1) \sigma^+(0,0)\right\rangle.
\end{eqnarray}
If the even fusion rules (\ref{efusionRules}) are satisfied  this implies
\[
C^{+}_{(\alpha_{rs},\delta), (\beta_2,0), (\beta_1,0)} =
- C^{-}_{(\alpha_{rs},\delta), (\beta_2,0), (\beta_1,0)}.
\]
It follows that
$C^{(-)}_{(\alpha_{rs},\delta), (\beta_2,0), (\beta_1,0)}  \neq 0 $ and therefore the corresponding form
has to vanish:
\[ \rho_{\scriptscriptstyle \rm NRe,o}^{(-)}(\chi_{rs}, w_2^+, w_{1}^+)=0.
 \]
Similarly, for the odd fusion rules (\ref{ofusionRules}) one gets
$C^{(+)}_{(\alpha_{rs},\delta), (\beta_2,0), (\beta_1,0)}  \neq 0 $ and
\[
\rho_{\scriptscriptstyle \rm NRe,o}^{(+)}(\chi_{rs}, w_2^+, w_{1}^+)=0.
 \]
An additional information on zeros of the forms
in question can be derived from the formula
$$
C^{(\pm)}_{(\alpha_{rs},\delta), (-\beta_2,0), (\beta_1,0)}
=
C^{(\mp)}_{(\alpha_{rs},\delta), (\beta_2,0), (\beta_1,0)},
$$
which is a simple consequence of (\ref{formula}). The
form $\rho_{\scriptscriptstyle \rm NRe,o}^{(+)}(\chi_{rs}, w_2^+, w_{1}^+)$
has to vanish for the even fusion rules (\ref{efusionRules})
and $\rho_{\scriptscriptstyle \rm NRe,o}^{(+)}(\chi_{rs}, w_2^+, w_{1}^+)$
 for the odd fusion rules (\ref{ofusionRules})
with the  opposite sign in front of $\beta_2$ in both cases.
The same reasoning  applies to the forms $\rho_{\scriptscriptstyle \rm RNp}^{(\pm)}(w_4^+, w_{3}^+,\chi_{rs} )$
as well.

The discussion above suggests the following definition of the fusion polynomials
in the Ramond sector:
\begin{equation}
P^{rs}_{c}\!\left[^{\pm \beta_2}_{\hspace{4pt}\beta_1}\right] \; =
\; \prod_{p=1-r}^{r-1} \prod_{q=1-s}^{s-1}
\left(\beta_1\mp \beta_2 + \frac{ p b+ q b^{-1}}{2\sqrt2}\right)
\prod_{p'=1-r}^{r-1} \prod_{q'=1-s}^{s-1}
\left(\beta_1\pm \beta_2 + \frac{ p' b+q' b^{-1}}{2\sqrt2}\right)
\end{equation}
where  $p,q,p',q'$ run with the step 2 and satisfy the conditions: $p+q -(r+s) \in 4{\mathbb Z} + 2$
 and $p'+q' -(r+s) \in 4{\mathbb Z}$. One easily checks that for
$rs \in 2 \mathbb{N}$,  $P^{rs}_{c}\!\left[^{ \pm \beta_2}_{\hspace{4pt}\beta_1}\right]$
are polynomials of degree $  \frac{rs}{2} $ in  $(\Delta_2 - \Delta_1)$,
and for  $rs \in 2 \mathbb{N} -1 $ -- of degree $\frac{rs-1}{2}$  in $(\Delta_2 - \Delta_1)$
 with the additional factor $(\beta_1 \mp \beta_2)$.
On the other hand
\begin{eqnarray*}
\rho^{(\pm)}_{\scriptscriptstyle \rm NRe}(L^n_{-1}\nu, w_2^+, w_1^+)\, &=& (\Delta + \Delta_2 - \Delta_1)_n ,
\\ [4pt]
\rho^{(\pm)}_{\scriptscriptstyle \rm NRo}(S_{- \frac12} L^n_{-1}\nu, w_2^+, w_1^+)
&=& {\rm e}^{- i{\pi\over 4}} (\beta_1 \mp \beta_2) (\Delta + \Delta_2 - \Delta_1)_n,
\\[4pt]
 \rho^{(\pm)}_{\scriptscriptstyle \rm RNe}( w_3^+, w_2^+, L^n_{-1}\nu)\,
 &=& (\Delta + \Delta_2 - \Delta_3)_n   ,
 \\ [4pt]
 \rho^{(\pm)}_{\scriptscriptstyle \rm RNo}(w_3^+, w_2^+, S_{- \frac12} L^n_{-1}\nu)
&=& - {\rm e}^{ i{\pi\over 4}} (\beta_3 \mp \beta_2) (\Delta + \Delta_2 - \Delta_3)_n,
 \end{eqnarray*}
where $(a)_n = \frac{\Gamma(a+n)}{\Gamma(a)}$ is the Pochhammer symbol.
  Taking into account our normalization condition for $\chi_{rs}$ one thus finally gets
\begin{equation}
\label{rhorsRN}
\begin{array}{lllllllll}
\begin{array}{rrrrrr}
  \rho_{\scriptscriptstyle \rm NRe}^{(\pm)}(\chi_{rs}, w_2^+, w_{1}^+)
  &=&
 \hspace{26pt}P^{rs}_{c}\!\left[^{\pm \beta_2}_{\hspace{4pt}\beta_1}\right],
 \\     [4pt]
  \rho_{\scriptscriptstyle \rm RNe}^{(\pm)}(w_3^+, w_{2}^+,\chi_{rs})
  &=&
 P^{rs}_{c}\!\left[^{\pm \beta_2}_{\hspace{4pt}\beta_3}\right] ,
\end{array}
 & {\rm for}& rs \in 2\mathbb{N},
 \\ [18pt]
\begin{array}{rrrrrr}
 \rho_{\scriptscriptstyle \rm NRo}^{(\pm)}(\chi_{rs}, w_2^+, w_{1}^+)
  &=&{\rm e}^{- i{\pi\over 4}} \, P^{rs}_{c}\!\left[^{\pm \beta_2}_{\hspace{4pt}\beta_1}\right],
  \\      [4pt]
\rho_{\scriptscriptstyle \rm RNo}^{(\pm)}(w_3^+, w_{2}^+,\chi_{rs})
  &=&  - {\rm e}^{ i{\pi\over 4}} \, P^{rs}_{c}\!\left[^{\pm \beta_2}_{\hspace{4pt}\beta_3}\right],
  \end{array}
& {\rm for}& rs \in 2\mathbb{N} -1.
\end{array}
\end{equation}

\section{Elliptic recurrence}

As in the bosonic and the NS cases the first step in a derivation of the elliptic recurrence
is to find the large $\Delta$ asymptotic of the conformal block.
The method of calculations proposed in \cite{Zamolodchikov:3} is based
on the observation that the full dependence of the first two terms in the large $\Delta$ expansion
on the variables $\Delta_i, c$
can be read off from the first two terms of the ${1\over \delta}$ expansion
of the classical block.
The essential point  of this approach  is
 the existence of the classical limit of conformal blocks.
In the present case this limit is to some extend justified by the
path integral representation
of the $N=1$ super-Liouville amplitudes
defined
 by the action:
\begin{equation}
\label{L:SLFT}
{\cal S}_{\rm\scriptscriptstyle SLFT}=
\int\!d^2z
\left(\frac{1}{2\pi}\left|\partial\phi\right|^2
+
\frac{1}{2\pi}\left(\psi\bar\partial\psi + \bar\psi\partial\bar\psi\right)
+
2i\mu b^2\bar\psi\psi {\rm e}^{b\phi}
+
2\pi b^2\mu^2{\rm e}^{2b\phi}\right).
\end{equation}
Within the functional approach the Ramond fields  are  represented
by vertex operators
(\ref{vertexop}).
Since the twist fields are light the fermionic sector does not contribute
to the classical limit
at all.
Strictly speaking  the path integral representation implies the classical limit
of the whole amplitude. Imposing extra restrictions on intermediate states one
may extend the argument to the sum of  bilinear block products with fixed parity and intermediate weight.
Considering various amplitudes with the same classical limit one may get
the information concerning individual terms. This leads to the assumption
that
 in the limit
$$
b\to 0,
\hskip 1cm
i b\beta_i \to  p_i,
\hskip 1cm
b^2\Delta_i \to \delta_i= p_i^2,
$$
the asymptotic behavior of each conformal block takes the form
\begin{eqnarray}
\label{claslim1}
\mathcal{F}_{\Delta}^{1}\!
        \left[^{\pm \beta_3 \; \pm \beta_2}_{\;\;\;\beta_4
\,\;\;\;\beta_1} \right]
         (z)
         &\sim &
       r_1 {\rm e}^{{1\over 2 b^2} f_{\delta}
\!\left[_{\delta_{4}\;\delta_{1}}^{\delta_{3}\;\delta_{2}}\right]
\!(z)},
\hskip 1cm
\mathcal{F}_{\Delta}^{1\over 2}\!
        \left[^{\pm \beta_3 \; \pm \beta_2}_{\;\;\;\beta_4
\,\;\;\;\beta_1} \right]
         (z)
         \;\sim \;
       r_{1\over 2} {\rm e}^{{1\over 2b^2} f_{\delta}
\!\left[_{\delta_{4}\;\delta_{1}}^{\delta_{3}\;\delta_{2}}\right]
\!(z)},
\end{eqnarray}
where $f_{\delta}
\!\left[_{\delta_{4}\;\delta_{1}}^{\delta_{3}\;\delta_{2}}\right]
\!(x)$ is the classical conformal block.
Analyzing the leading powers of $\Delta$ in the  forms
$\rho^{(\pm)}_{\scriptscriptstyle \rm NR },\rho^{(\pm)}_{\scriptscriptstyle \rm RN }$
\begin{eqnarray*}
\nonumber
\rho^{(\pm)}_{\scriptscriptstyle \rm NR,e }(\nu_{\Delta,KM}, w_2^+, w_1^+) \, \propto \,
\Delta^{|K|+|M|}+\dots \ ,
\quad
\rho^{(\pm)}_{\scriptscriptstyle \rm NR,o }(\nu_{\Delta,KM}, w_2^+, w_1^+) &\propto &
\beta_i \Delta^{|K|+|M|-{1\over 2}}+\dots \ ,
\\
\rho^{(\pm)}_{\scriptscriptstyle \rm RN,e}(w_3^+, w_2^+, \nu_{\Delta,KM})  \, \propto \,
  \Delta^{|K|+|M|}+\dots \ ,
\quad
\rho^{(\pm)}_{\scriptscriptstyle \rm RN,o}(w_3^+, w_2^+, \nu_{\Delta,KM})  &\propto &
 \beta_i \Delta^{|K|+|M|-{1\over 2}}+\dots \ ,
\end{eqnarray*}
one can show that the coefficients
$r_1, r_{1\over 2}$
are independent of $b$ and
 $$
r_1\propto {\rm const},
\hskip 1cm
r_{1\over 2} \propto {1\over \delta},
$$
as functions of $\delta$.

Once the classical limits are known one can follow Zamolodchikov's
derivation in order to find the large $\Delta$ behavior. In the present case it yields:

\begin{eqnarray}
\label{asymptoticG}
\nonumber
&& \hskip -1cm \ln
\mathcal{F}_{\Delta}^{1}\!
        \left[^{\pm \beta_3 \; \pm \beta_2}_{\;\;\;\beta_4
\,\;\;\;\beta_1} \right]
         (z)
         =
\pi \tau \left( \Delta - \frac{c}{24} \right)
+ \left( \frac{c}{8} - \Delta_1 - \Delta_2 - \Delta_3 - \Delta_4\right) \, \ln{K^2(z)} \\
 [6pt] &&
+ \left( \frac{c}{24} - \Delta_2 - \Delta_3\right)  \, \ln(1-z)
+  \left( \frac{c}{24} - \Delta_1 - \ \Delta_2\right)  \, \ln(z) + f^{\pm\pm}(z) +{\cal O}\left({1\over \Delta}\right),
\\
\nonumber
\label{asymptoticGo}
&& \hskip -1cm \ln
\mathcal{F}_{\Delta}^{1\over 2}\!
        \left[^{\pm \beta_3 \; \pm \beta_2}_{\;\;\;\beta_4
\,\;\;\;\beta_1} \right]
         (z)
   =
 -\ln \Delta
+\pi \tau \left( \Delta - \frac{c}{24} \right)
+ \left( \frac{c}{8} - \Delta_1 - \Delta_2 - \Delta_3 - \Delta_4\right) \, \ln{K^2(z)} \\ [6pt]
&&
+ \left( \frac{c}{24} - \Delta_2 - \Delta_3\right)  \, \ln(1-z)
+  \left( \frac{c}{24} - \Delta_1 - \ \Delta_2\right)  \, \ln(z) +{\cal O}\left({1\over \Delta}\right),
\end{eqnarray}
where
$$
\tau=i{K(1-z)\over K(z)}
$$
and  $f^{\pm\pm}(z)$ are functions of $z$ specific for each type of block and independent of $\Delta_i$ and $c$.
 The large $\Delta$ asymptotic suggests the following form of superconformal blocks:
 \begin{eqnarray}
\label{Hblock}
  \mathcal{F}^{1, \frac12}_{\Delta}\!
\left[^{\pm \beta_3 \; \pm \beta_2}_{\;\;\;\beta_4
\,\;\;\;\beta_1} \right] (z)
 & =&
(16q)^{\Delta - \frac{c-3/2}{24}}\ z^{\frac{c-3/2}{24} - \Delta_1 -\Delta_2} \
 (1- z)^{\frac{c-3/2}{24} - \Delta_2 - \Delta_3}\
\\
\nonumber
 & \times &\theta_3^{\frac{c - 3/2}{2}
- 4 (\Delta_1 +\Delta_2 +\Delta_3 + \Delta_4) }  \
 \mathcal{H}^{1, \frac12}_{\Delta}\! \left[^{\pm \beta_3 \; \pm \beta_2}_{\;\;\;\beta_4
\,\;\;\;\beta_1} \right]\!(q),
\end{eqnarray}
where
$
q=\exp(i\pi \tau)\ .
$
The
elliptic blocks $\mathcal{H}^{1, \frac12}_{\Delta}\! \left[^{\pm \beta_3 \; \pm \beta_2}_{\;\;\;\beta_4
\,\;\;\;\beta_1} \right]\!(q)$ have the same analytic structure as
superconformal ones:
\begin{eqnarray*}
 \mathcal{H}^{1,\frac12}_{\Delta}\! \left[^{\pm \beta_3 \; \pm \beta_2}_{\;\;\;\beta_4
\,\;\;\;\beta_1} \right]\!(q)&=&
  g^{1,\frac12}_{\pm\pm}(q)
+ \sum_{m,n} \frac{h^{1,\frac12 }_{mn}\left[^{\pm \beta_3 \; \pm \beta_2}_{\;\;\;\beta_4
\,\;\;\;\beta_1} \right](q)}{\Delta - \Delta_{mn}}.
\end{eqnarray*}
The functions  $ g^{1,\frac12}_{\pm\pm}(q)$
depend on the block type and are independent
of the external momenta $\beta_i$ and the central charge $c$. Since they are non-singular in  $\Delta$, it follows from (\ref{asymptoticG}), (\ref{asymptoticGo}) that
 $ g^{1}_{\pm\pm}(q)$ are directly related to the functions
$f^{\pm\pm}(z)$ and $ g^{\frac12}_{\pm\pm}(q)=0$ .

The functions $ g^{1,\frac12}_{\pm\pm}(q)$
depend on the block type and are independent
of the external momenta $\beta_i$ and the central charge $c$.
They have no singularities in $\Delta$ and are directly related to the functions
$f^{\pm\pm}(z)$ in (\ref{asymptoticG}).
The analytic form of these functions can be read off from the  $\hat c = 1$
elliptic blocks with $\Delta_i= \Delta_0 = \frac{1}{16}$.
The explicit formula for this blocks can be obtained using the techniques
of the chiral superscalar model \cite{Hadasz:2007ns}.
The chiral correlation function projected on the intermediate $\Delta$ NS module
of the fields $\sigma_0$ corresponding to the lowest Ramond state in the bosonic sector and
the NS vacuum state in the fermionic sector takes the form:
\begin{equation}
\label{specblock}
\big\langle\sigma_0\big|
\sigma_0(1)\big|_{\Delta}\sigma_0(z)
\big|\sigma_0\big\rangle
=
|\big\langle\nu_\Delta \big|
\sigma_0(1)
\big|\sigma_0\big\rangle|^2\
(16q)^{\Delta}\,
\left[z(1-z)\right]^{-\frac18}\,
\theta_3(q)^{-1}.
\end{equation}
On the other hand the $b\to i$, $\beta_i\to 0$ limit of each type of general even block
is regular for generic values of $\Delta$ and yields
$$
\lim_{\beta\to 0}\lim_{b\to i}
 \mathcal{F}^{1}_{\Delta}\!
\left[^{\pm \beta \; \pm \beta}_{\;\;\;\beta
\,\;\;\;\beta} \right] (z)
=(16q)^{\Delta}\,
\left[z(1-z)\right]^{-\frac18}\,
\theta_3(q)^{-1}
g^{\pm\pm}(q)\ .
$$
Comparing with (\ref{specblock}) one gets
$$
g^1_{\pm\pm}(q)=1\ .
$$
Collecting the results one can write the elliptic recurrence in the Ramond sector in the following form
\begin{eqnarray*}
\nonumber
 \mathcal{H}^{e/o}_{\Delta}\! \left[^{\pm \beta_3 \; \pm \beta_2}_{ \;\;\; \beta_4 \; \;\;\;\beta_1} \right]\!(q)
=
 g^{1,\frac12}_{\pm\pm}(q)
 \! &+&\!
 \begin{array}[t]{c}
{\displaystyle\sum} \\[-5pt]
{\scriptscriptstyle
r,s>0}
\\[-7pt]
{\scriptscriptstyle
r,s\in 2{\mathbb N}
}
\end{array}
 \hspace{-7pt} (16q)^{\frac{rs}{2}}
 \frac{ A_{rs}(c)
P^{rs}_{c}\!\left[^{\pm\beta_3}_{\;\;\; \beta_4}\right]
P^{rs}_{c}\!\left[^{\pm\beta_2}_{\;\;\; \beta_1} \right]
}{\Delta - \Delta_{rs}} \,
\mathcal{H}^{e/o}_{\Delta_{rs}+\frac{rs}{2}}\!
\left[^{\pm \beta_3 \; \pm \beta_2}_{ \;\;\; \beta_4 \; \;\;\;\beta_1} \right]\!(q)
 \\[-6pt]
 \\[-6pt] \nonumber
\! &-&\! \hspace{-9pt} \begin{array}[t]{c}
{\displaystyle\sum} \\[-5pt]
{\scriptscriptstyle
r,s>0}
\\[-7pt]
{\scriptscriptstyle
r,s\in 2{\mathbb N}+1
}
\end{array}
\hspace{-7pt} (16q)^{\frac{rs}{2}}
\frac{   A_{rs}(c) \,
P^{rs}_{c}\!\left[^{\pm\beta_3}_{\;\;\; \beta_4}\right]
P^{rs}_{c}\!\left[^{\pm\beta_2}_{\;\;\; \beta_1} \right]}{\Delta - \Delta_{rs}} \,
\mathcal{H}^{o/e}_{\Delta_{rs}+\frac{rs}{2}}\!
\left[^{\mp \beta_3 \; \mp \beta_2}_{ \;\;\; \beta_4 \; \;\;\;\beta_1} \right]\!(q).
\end{eqnarray*}

\setcounter{equation}{0}
\section*{Acknowledgements}
The work  was supported by the Polish State Research
Committee (KBN) grant no. N~N202 0859 33.

\noindent
P.S. is grateful to the faculty of the Institute of Theoretical Physics, University of Wroc{\l}aw for
the hospitality.

\end{document}